\documentclass [prb,onecolumn,notitlepage,superscriptaddress,floatfix] {revtex4-1}
\usepackage{amsmath}
\usepackage{graphicx}
\usepackage{lmodern}
\usepackage{amsmath}
\usepackage{comment}

\usepackage{color}
\usepackage{amssymb}
\newcommand{\beq} {\begin{equation}}
\newcommand{\eeq} {\end{equation}}
\newcommand{\bea} {\begin{eqnarray}}
\newcommand{\eea} {\end{eqnarray}}
\newcommand{\be} {\begin{equation}}
\newcommand{\ee} {\end{equation}}
\renewcommand{\(}{\left(}
\renewcommand{\)}{\right)}
\renewcommand{\[}{\left[}
\renewcommand{\]}{\right]}

\DeclareMathOperator{\sgn}{sgn}
\DeclareMathOperator{\Tr}{Tr}

\begin{document}

\title {Interplay between pair-density-wave and charge-density-wave orders in underdoped cuprates}
\author{Yuxuan Wang}
\affiliation{Department of Physics, University of Wisconsin, Madison, WI 53706, USA}
\author{Daniel F. Agterberg}
\affiliation{Department of Physics, University of Wisconsin-Milwaukee, Milwaukee, Wisconsin 53211, USA}
\author{Andrey Chubukov}
\affiliation{Department of Physics, University of Wisconsin, Madison, WI 53706, USA}
\affiliation{William I. Fine Theoretical Physics Institute,
and School of Physics and Astronomy,
University of Minnesota, Minneapolis, MN 55455, USA}
\date{December 15, 2014}

\begin{abstract}
We analyze the interplay between charge-density-wave (CDW)  and pair-density-wave (PDW) orders within the spin-fermion model for the cuprates. We
 specifically consider  CDW order with transferred momenta $(\pm Q,0)$/$(0,\pm Q)$, as seen in experiments on the cuprates,
  and PDW order with total momenta $(0,\pm Q)/(\pm Q,0)$.
Both orders  have been proposed to explain the pseudogap phase in the cuprates.
 We show that both emerge in the spin-fermion model near the onset of antiferromagnetism.  Each order parameter is constructed out of  pairs of fermions in ``hot" regions on the Fermi surface, breaks ${\rm U}(1)$ translational symmetry, and changes sign when the  momenta of the fermions change by $(\pi,\pi)$. We further show that the two orders
 are nearly degenerate due to an approximate ${\rm SU}(2)$ particle-hole symmetry of the model. This near degeneracy is similar in origin to that relating
  conventional $d$-wave superconducting order and bond charge order with momentum $(Q, \pm Q)$.
 The ${\rm SU}(2)$ symmetry becomes exact if one neglects the curvature of the Fermi surface in hot regions, in which case ${\rm U}(1)$ CDW and PDW order parameters
 become components of
  an ${\rm SO}(4)$-symmetric PDW/CDW ``super-vector".
   We develop a Ginzburg-Landau theory for four PDW/CDW order parameters and find two possible ground states: a ``stripe" state in which
  both CDW and PDW orders  develop with either $(\pm Q,0)$ or $(0,\pm Q)$,
  and a ``checkerboard" state, where each order can develop with $(\pm Q,0)$ and $(0,\pm Q)$.
     We
   show that the ${\rm SO}(4)$ symmetry between CDW and PDW can be broken by two separate effects. One is the inclusion of Fermi surface curvature, which  selects a PDW order immediately below the instability temperature. Another is the overlap between different hot regions, which
   favors CDW order at low temperatures. For the stripe state, we show that  the  competition between the two effects
    gives rise to a first-order transition from PDW to CDW inside the ordered state.  We also argue that beyond mean-field theory, the onset temperature for CDW order is additionally enhanced due to feedback from a preemptive breaking of ${\mathbb Z}_2$ time-reversal symmetry.
    We discuss the ground state properties of a pure PDW state and a pure CDW state, and show in particular that the PDW checkerboard state yields a vortex-anti-vortex lattice.
      For the checkerboard state, we considered a situation  when both CDW and PDW orders are present at low $T$ and show that
       at small but finite Fermi surface curvature
       the presence of both condensates  induces a  long sought chiral $s+id_{xy}$ superconductivity.
\end{abstract}
\maketitle

\section{Introduction}
Understanding the nature of the pseudogap in the cuprates is a necessary step in solving the puzzle of high-$T_c$ superconductivity.
  At present, most researchers agree that the pseudogap is more than just a precursor to superconductivity~\cite{emery} or a result of a strong
   fermionic incoherence due to the interaction with some near-featureless overdamped boson (e.g, a paramagnon~\cite{acs}).
    There is, however, no consensus on the primary origin of the pseudogap behavior.  Some researchers cite  experimental indications for static charge order~\cite{ybco,ybco_1,X-ray, X-ray_1,davis_1,mark_last} in at least part of the pseudogap region
   as evidence that  pseudogap behavior is associated with the development of
      a new electronic order.
       This order can be
       either an incommensurate  charge-density-wave  order (CDW)~\cite{efetov,charge}
       defined as $\rho_{\bf k}^{\bf Q} \propto \langle c_\alpha^\dagger({\bf k+Q}/2) c_\beta({\bf k-Q}/2)\rangle \delta_{\alpha\beta}$
       or an incommensurate pair-density-wave order
       (PDW)~\cite{patrick,agterberg,kivelson},
       $\varphi_{\bf Q}^{\bf k}\propto \langle c_\alpha({\bf k-Q}/2) c_\beta(-{\bf k-Q}/2)\rangle (i\sigma^y_{\alpha\beta})$. The latter is
       a superconducting (SC) order with a finite Cooper pair momentum
        ${\bf Q}$, much like FFLO~\cite{FF,LO} state but at zero magnetic field.
     Each of these orders competes with conventional $d$-wave superconductivity  and the competition pushes the $d$-wave $T_c$ down in underdoped cuprates.
     Others argue that  CDW or PDW orders are secondary effects in the pseudogap phase, and the primary reason for the pseudogap behavior is a localization of an electron due to close proximity of the insulating Mott state at half-filling~\cite{anderson,lee,rice,millis,tremblay_1,ph_ph,lee_senthil}.
     The competing order and Mott  scenarios for the pseudogap are not necessarily orthogonal as there is little doubt that precursors to Mott physics  do develop near half-filling
      and can, in principle, enhance CDW and/or PDW correlations~\cite{lee_senthil}.
     However, if a competing order develops at a higher doping,
     it may give also rise to pseudogap behavior while the system still remains a metal. In this case, the quasiparticle residue is reduced only at low frequencies and not over the whole bandwidth, as is the case for precursors to Mott physics.
     To analyze this scenario it becomes necessary to study the behavior of an
     electronic system
 in a parameter range where competing orders develop but electrons remain delocalized (itinerant)~\cite{acs,ms}.

 The CDW  scenario has been proposed for the cuprates a while ago~\cite{grilli, kontani, castellani, extra_gr},  but has recently gained momentum~\cite{ms,16,17,laplaca,19,20,21,efetov,23,24,25,26,27,29,tsvelik,31,atkinson,33,debanjan,35}
  due to strong experimental evidence for charge order in the underdoped cuprates. On the theory side,  Metlitski and Sachdev~\cite{ms} (MS)
   considered a model of fermions interacting by exchanging $(\pi,\pi)$  spin fluctuations (a spin-fermion model) and
  have shown  that magnetically mediated interaction between hot fermions, known to give rise to $d$-wave superconductivity, also gives rise to
  CDW order with a $d$-wave form factor. A $d$-wave CDW order is often called bond order (BO) as in real space it affects bond charge density $\langle c^\dagger_{r} c_{r+a}\rangle$ but does not affect on-site density $\langle c^\dagger_{r} c_{r}\rangle$.
     MS  showed that, not only charge order emerges in the spin-fermion model, but the critical temperature for this order is
    identical to that for superconductivity if one neglects the curvature of the Fermi surface in hot regions. Subsequent research along these lines
      led to a proposal~\cite{efetov} that the pseudogap may be due to the fact that, over a wide range of $T$, the system
   cannot decide between near-degenerate bond charge order and SC order.
   This proposal is appealing but, at least at a face value, is inconsistent with the experiments because the  bond charge order has momentum along one of Brillouin zone diagonals ($Q, \pm Q$), while the charge order  detected in the experiments is along $x$ or $y$ directions in
  momentum space, i.e.,
   it has momentum $(Q,0)$ or $(0,Q)$.

   CDW order with momentum $(Q,0)/(0,Q)$ also emerges from the spin-fluctuation scenario~\cite{charge}.
    The form-factor for this order has both $d$-wave and $s$-wave components and, in real space, gives rise to variations of both on-site and bond charge density.
    At the mean-field level  the instability temperature for $(Q,0)/(0,Q)$ order is smaller than that for $(Q,Q)$ order.
      There exist two proposals how one can  obtain  CDW order with momentum $(Q,0)/(0,Q)$ as a leading instability.
      One proposal is to assume that some precursors to Mott physics~\cite{debanjan} (or Heisenberg antiferromagnetism~\cite{31,atkinson,33})
       develop before charge fluctuations get soft.
      In this situation, charge order develops from an already reconstructed fermionic dispersion, and  calculations show~\cite{31,atkinson,33,debanjan}
       that already within the mean-field (Hartree-Fock) approximation $(Q,0)/(0,Q)$,  order develops at a higher $T$ than than $(Q,Q)$ order. Another proposal is to go beyond the mean-field approximation and consider fluctuation effects. Along these lines,
      it has been
      shown~\cite{charge,tsvelik} that  CDW order with momenta along $x$ or $y$ direction additionally breaks $C_4$ lattice rotational symmetry down to $C_2$
       (it is either along $x$ or along $y$)
       and  also breaks time-reversal and mirror symmetries. Both symmetry breakings are consistent with experiment: the breaking of $C_4$ symmetry has been detected in STM
       resistivity, STM, and thermo-electric coefficient measurements~\cite{ando,davis_1,taillefer_last}, and the breaking of time-reversal and mirror symmetries gives rise to a non-zero Kerr effect, as observed  by the Kapitulnik group~\cite{YBCO_kerr, LBCO_kerr}.
    The authors of Refs.\ \onlinecite{charge,tsvelik} demonstrated that the discrete ${\mathbb Z}_2$ symmetries get broken at a higher $T$ than a  temperature at which an
     incommensurate CDW order emerges, and this effect pushes
   the actual onset temperature of CDW order up.  No such enhancement occurs for $(Q,Q)$ order, which does not break any discrete symmetry.
      In this paper we assume that, for one reason or another, CDW order with $(Q,0)/(0,Q)$
       wins over $(Q,Q)$ order and only consider CDW order with $(Q,0)/(0,Q)$ which below we just call CDW.

      In a separate line of development, the authors of Ref.\ \onlinecite{kivelson} argued that the data on La$_{2-x}$Ba$_x$CuO$_4$ show evidence for PDW order, along with
       SDW and CDW orders, and claimed that anomalous SC fluctuations, possibly associated with  short-range PDW, exist also in
       YBa$_2$Cu$_3$O$_{6+y}$ \cite{ber09,agterberg_2}.   PDW order has been also shown to emerge in a doped Mott insulator~\cite{lee_senthil,corboz}.
      P.\ A. Lee argued~\cite{patrick} that ARPES data on the change of the fermionic dispersion in the pseudogap phase can be best described
      if one assumes that the system develops PDW order rather than CDW order.  One of us further demonstrated~\cite{agterberg} that PDW order can, by itself, give rise to breaking of
       $C_4$ and time-reversal symmetries.  This suggests that PDW order is another viable candidate for a competing order parameter.

        In this paper we show that, in a magnetic scenario,  both PDW and CDW orders develop from magnetically mediated interaction between hot fermions.  We  show that when the curvature of the Fermi surface in hot regions is neglected,
        PDW and CDW orders are related by ${\rm SU}(2)$ symmetry, just like $d$-wave superconductivity and bond charge order,
         and the critical temperatures for PDW and CDW  orders are identical.
      The degeneracy between PDW and CDW orders in the hot spot model has been previously noticed Ref.\ \onlinecite{pepin} and our results on the degeneracy
       between PDW and CDW orders agree with theirs.   Ref.\ \onlinecite{pepin} then focused on the interplay between CDW/PDW order and SC/BO within the hot spot model
        and explored that the fact that mean-field transition temperature for SC/BO is higher than that for CDW/PDW.  We focus on CDW/PDW subset and
         analyse the type of CDW/PDW order (e,g., stripe or checkerboard)
        and the selection of the order in both stripe and checkerboard phases by  going beyond the hot spot model and beyond mean-field approximation.
          We address the interplay between CDW/PDW and d-wave SC order in a separate paper ~\cite{coex}.

             The paper is organized as follows. In Sec.\ II we obtain and compare critical temperatures for CDW and PDW instabilities  within the spin-fermion model
         with a linear dispersion near the hot spots. We first verify that $T_{\rm CDW}$ and $T_{\rm PDW}$ are identical through explicit calculation, and then demonstrate that this is a direct consequence of ${\rm SU}(2)$ symmetry of the
          underlying fermionic
          model. In this Section we also discuss the structure of CDW and PDW order parameters
          $\rho_{\bf k}^{\bf Q}$ and $\varphi_{\bf Q}^{\bf k}$ as functions of momentum ${\bf k}$.
          In particular, we show that both orders must change sign under the change of
          ${\bf k}$  by $(\pi,\pi)$.
         In Sec.\ III we derive the GL effective action for the coupled CDW and PDW order parameters made out of fermions concentrated near hot spots (which are Fermi-surface points for which $k_F$ and $k_F +  (\pi,\pi)$ are both on the Fermi surface). In Sec.\ IV we analyze the structure of the ground state configuration and show that two different states -- a stripe state and a checkerboard state --  are possible, depending on system parameters.
            In Sec.\ \ref{sec:5} we discuss the system behavior when Fermi surface curvature is not neglected and order parameters are allowed to couple to fermions away from hot spots.  We first show in Sec.\ \ref{5A} that a finite curvature breaks the degeneracy between CDW and PDW orders and that, at a mean-field level, $T_{\rm PDW}$ becomes larger than $T_{\rm CDW}$. We discuss the properties of a pure PDW state in Sec.\ \ref{5B}.
            In particular, we show that the
  checkerboard PDW ground state can be viewed in  real space as  a vortex anti-vortex lattice. We then show in Sec.\ \ref{5C}  that we get an opposite effect, that is $T_{\rm CDW}$ becomes larger than $T_{\rm PDW}$,
             when we allow CDW and PDW order parameters  couple to fermions away from hot spots.  This gives rise to extra terms in the GL action which favor the CDW state. We briefly discuss the properties of a pure CDW state in Sec.\ \ref{5D}. In Sec.\ \ref{5E} we combine the two effects and show that the system undergoes a first-order transition from PDW to CDW inside the ordered phase.  In Sec.\ \ref{5F} we extend the considerations beyond mean-field and show that the onset temperature for CDW order $T_{\rm CDW}$ is additionally enhanced due to feedback effect from a preemptive breaking of ${\mathbb Z}_2$ time-reversal symmetry,
               and can potentially become larger than $T_{\rm PDW}$  even if at a mean-field level $T_{\rm PDW}$ was larger.
           In Sec.\ VI we discuss the development of a secondary homogeneous  SC order, first in general terms in Sec.\ \ref{6A} and then specifically for our model in Sec.\ \ref{6B}. We  show that in checkerboard states in which both CDW an PDW orders are present,
  this secondary order parameter
   at small but finite Fermi surface (FS) curvature has $s+id_{xy}$ symmetry.
   We extend the analysis of secondary homogeneous SC order beyond the leading order in the curvature and find that, in general,  $d_{x^2-y^2}$ and $A_{2g}$ SC orders are also induced.
             The amplitude of  $d_{x^2-y^2}$ order is additionally enhanced because the interaction in this channel is attractive on its own.
                 We present our conclusions in Sec.\ VII
                 and briefly discuss the relation of our results to the physics of the pseudogap phase in the cuprates.

\section{PDW and CDW instabilities in the spin-fermion model with linear dispersion}
\begin{figure}
\centerline{\parbox{0.35\columnwidth}{\includegraphics[width=0.4\columnwidth]{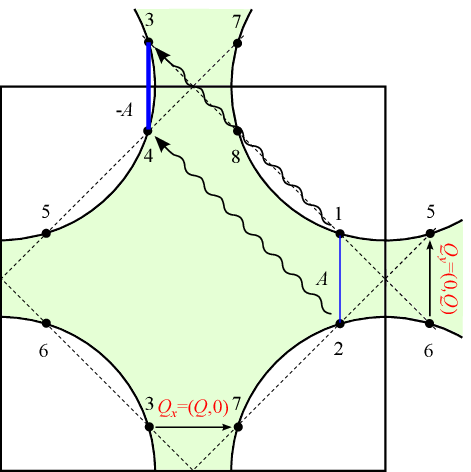}\\(a)}~~~~
\parbox{0.5\columnwidth}{\includegraphics[width=0.5\columnwidth]{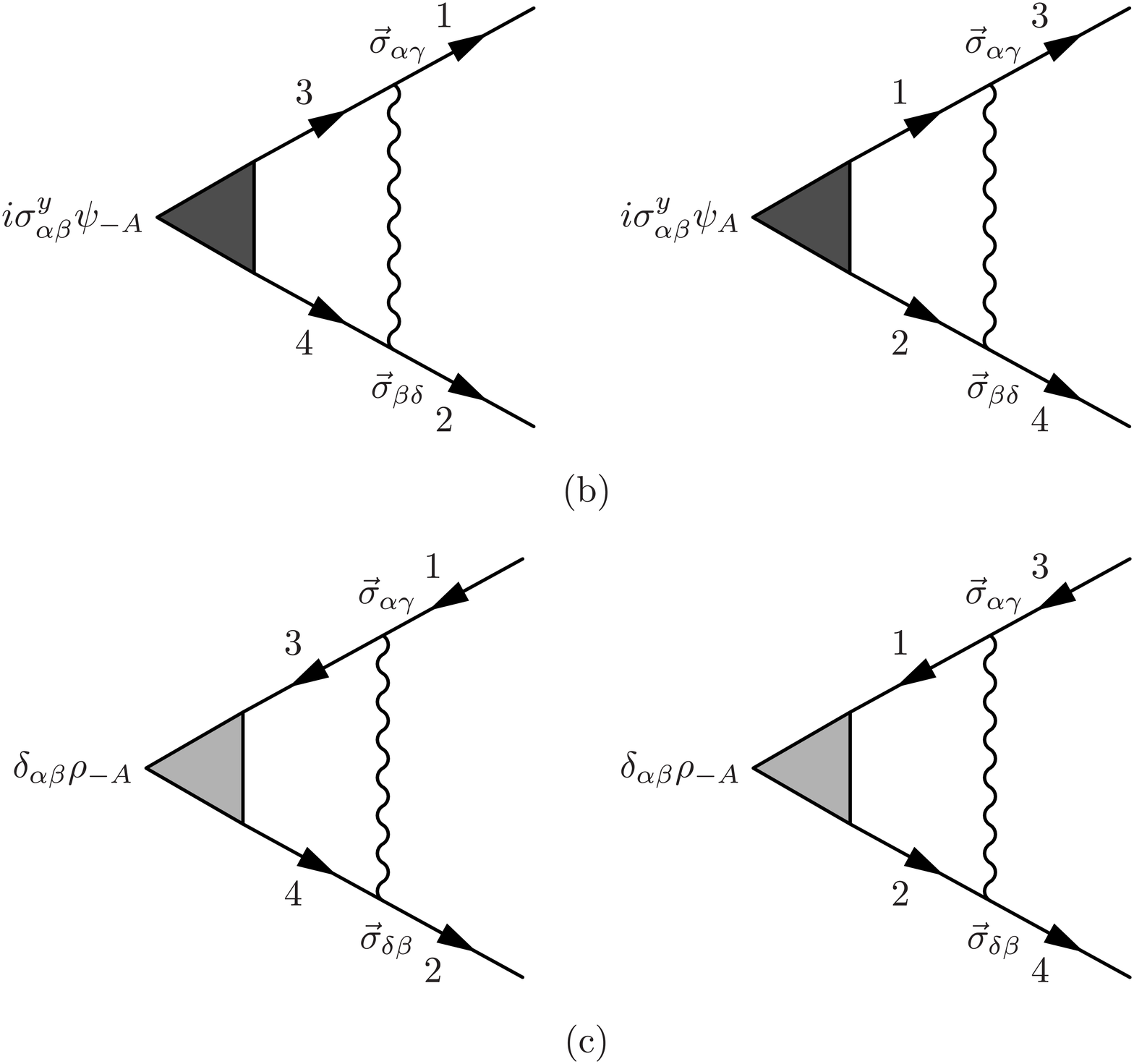}}}
\caption{Panel (a): The Fermi surface, Brillouin zone and magnetic Brillouin zone (dashed line). Hot spots are defined as interSections of the FS with magnetic
Brillouin zone. The hot spot pairs 1-2 and 3-4 denotes the CDW/PDW pairings we consider. They are coupled through the antiferromagnetic exchange interaction
peaked at momentum $(\pi, \pi)$, as shown by the dashed arrows.
Panel (b) and (c): Diagrammatic representation of CDW [Panel (b)] and PDW [Panel (c)] instabilities between hot spots (1,2) and (3,4).}
\label{cpdw}
\end{figure}
In this Section, we compare PDW and CDW instabilities in the spin-fermion model.
 This model has been intensively investigated in studies of non-Fermi-liquid physics~\cite{acs}, $d$-wave superconductivity~\cite{acf,ms,wang,wang_el}, charge-density-wave order~\cite{ms,efetov,charge}, and symmetry breaking in the pseudogap region~\cite{charge}. The model describes low-energy fermions with the Fermi surface shown in Fig.\ \ref{cpdw}(a) and with an effective four-fermion interaction mediated by soft spin collective excitations peaked at momentum transfer $(\pi,\pi)$. We focus on ``hot" regions on the Fermi surface, for which shifting the momentum by $(\pi,\pi)$ keeps a fermion near the Fermi surface. We show these hot spots in Fig.\ \ref{cpdw}(a) and label them as 1-8. Near a given hot spot $i$ we expand the fermionic dispersion as $\epsilon_{i,\tilde k}=v_{F,i}(\tilde k_{i,\perp}+\kappa\tilde k_{i,\|}^2/k_F)$, where ${\bf v}_{F,i}$ is the Fermi velocity at a given hot spot, $\tilde k_{i,\perp}$ and $\tilde k_{i,\|}$ are the deviations from the hot spot perpendicular to and along the Fermi surface, and dimensionless $\kappa$ specifies the curvature of the Fermi surface at the hot spot. In this and the next two Sections we linearize the fermionic dispersion, i.e., neglect $\kappa$.  We will discuss the effect of $\kappa$ in Section V.
We define the Fermi velocity at hot spot 1 as ${\bf v}_{F,1}=(v_x, v_y)$
(the velocities at other hot spots follow from symmetry),
 and define the momentum difference between hot spots 1 and 2 (5 and 6) as $Q_y=(0,Q)$, and the momentum difference between hot spots 3 and 7 as $Q_x=(Q,0)$.

The action of the spin-fermion model can be written as
\begin{align}
\mathcal{S}=&\sum_{i,\alpha}\int d\tilde k c^\dagger_{i\alpha}(\tilde k)(-i\omega_m+\epsilon_{i,\tilde k})c_{i\alpha}(\tilde k)+\frac{1}{2}\int dq \chi_0^{-1}(q)\vec\phi(q)\cdot\vec\phi(-q)\nonumber\\
&+g\sum_{\substack{i=1,2,5,6;\\\alpha\beta}}\int d\tilde kd\tilde k'\[c_{i\alpha}^\dagger(\tilde k)c_{i+2,\beta}(\tilde k')+c_{i+2,\alpha}^\dagger(\tilde k)c_{i,\beta}(\tilde k')\]\vec\sigma_{\alpha\beta}\cdot\vec\phi(\tilde k-\tilde k')
\label{sf}
\end{align}
where $c_{i\alpha}$ is fermion field with $i$ labeling hot spots and $\alpha,\beta$ labeling spin. Hot spots $i$ and $i+2$ are separated by $(\pi,\pi)$. The vector field $\vec \phi$ is the spin collective excitation. We have used shorthands $\tilde k=(\omega_m,{\bf \tilde k})$, $q=(\Omega_m,{\bf q})$, and $\omega_m(\Omega_m)$ are fermionic (bosonic) Matsubara frequencies. The bosonic momentum ${\bf q}$ is measured as the deviation from the antiferromagnetic momentum $(\pi, \pi)$, and the fermionic momentum ${\bf \tilde  k}$ is measured as the deviation from the corresponding hot spot. The static spin susceptibility in (\ref{sf}) has Ornstein-Zernike form
$\chi_0(q)={\chi_0}/{({\bf q}^2+\xi^{-2})}$.

The spin-fermion interaction gives rise to bosonic and fermionic self-energies. The bosonic self-energy is the Landau damping $\Pi(\Omega)=\gamma|\Omega|$, where $\gamma=4\bar g/(\pi|v_y^2-v_x^2|)$ and $\bar g=g^2 \chi_0$ (Ref.\ \onlinecite{acs}).
  The fermionic self energy is most singular at hot spots. When $\xi^{-1}=0$, $\Sigma(\omega_m,{\bf k}_{hs})=\Sigma(\omega_m)$ has a
 non-Fermi-liquid form, $\Sigma(\omega_m)=iA\sgn(\omega_m)\sqrt{\omega_0|\omega_m|} $, where
 $\omega_0=9\bar g/(16\pi)\times[(v_y^2-v_x^2)/v_F^2]$ and $A =1$ in the large $N$ approximation ($N$ is the number of fermionic flavors) and $A\approx2/3$ in a self-consistent rainbow approximation for the physical case of $N=1$ (see Eq.\ (9) in  Ref.\ \onlinecite{charge}).

   Following earlier work~\cite{acs, ms, efetov}, we assume that the coupling $\bar g$ is small compared to the Fermi energy $E_F=v_Fk_F/2$ and study instabilities which occur at energies well below $E_F$ and at $\xi^{-1} \geq 0$, i.e., before the system becomes magnetically ordered. Known instabilities include $d$-wave superconductivity~\cite{acf,ms,wang} and charge orders of momentum $(Q,Q)$ (bond charge orders, Refs.\ \onlinecite{ms,efetov}) and $(Q,0)/(0,Q)$ (CDW order,
   Refs.\ \onlinecite{charge,debanjan}).  We show that there exists another instability towards a SC order that breaks translational symmetry --
    a PDW order~\cite{agterberg, patrick, fradkin,kivelson}.

   In the next subsection we show that PDW and CDW orders are degenerate by explicitly studying linear self-consistency equations for the PDW and CDW condensates.
    Then we show that such a degeneracy is in fact a direct consequence of ${\rm SU}(2)$ particle-hole symmetry.

   \subsection{
    Ladder equations for CDW and PDW condensates}

 We define CDW and PDW condensates as
  \begin{align}
 i\sigma_{\alpha\beta}\varphi_{\bf Q}^{\bf k}\propto \langle c_\alpha({\bf k-Q}/2) c_\beta(-{\bf k-Q}/2)\rangle{\rm,~~~and~~~}\delta_{\alpha\beta}\rho_{\bf k}^{\bf Q} \propto \langle c_\alpha^\dagger({\bf k+Q}/2) c_\beta({\bf k-Q}/2)\rangle,
 \label{equation2}
 \end{align}
 where ${\bf Q}=\pm Q_y=(0,\pm Q)$ or $\pm Q_x=(\pm Q,0)$.
   It is trivial to verify that both $\varphi_{\bf Q}^{\bf k}$ and $\rho_{\bf k}^{\bf Q}$ carry momentum $\bf Q$. We note in this regard
   that  PDW condensate $\varphi_{\bf Q}^{\bf k}$ and CDW condensate $\rho_{\bf k'}^{\bf Q'}$ formed between {\it the same} fermions have  different momenta.
    Indeed, matching fermionic momenta for $\varphi_{\bf Q}^{\bf k}$ and $\rho_{\bf k'}^{\bf Q'}$, we find ${\bf Q}=-2{\bf k'}$ and ${\bf Q'}=2{\bf k}$.
      If the two fermions are in the vicinity of hot spots 1 and 2,  ${\bf k}=(0,Q/2)$ and ${\bf k'}=(Q/2,0)$. Then
      ${\bf Q}=(Q,0)=Q_x$ and ${\bf Q'}=(0,Q)=Q_y$. Therefore, the PDW and CDW condensates formed by same pair of hot fermions actually carry {\it orthogonal} momenta.

 The CDW and PDW order parameters couple to bilinear fermionic operators as
\begin{align}
S_{\rm int} = i\sigma_{\alpha\beta}\varphi_Q^kc^\dagger_{\alpha}(k-Q/2)c^\dagger_{\beta}(-k-Q/2)+\delta_{\alpha\beta}\rho_k^Q c_\alpha^\dagger(k-Q/2)c_\beta(k+Q/2)+h.c.
\label{25}
\end{align}
These couplings are renormalized by four-fermion interactions. To analyze PDW and CDW orders, one needs to solve self-consistent equations for $\rho$ and $\varphi$. For definiteness, we focus on hot spots (1, 2) and (3, 4) in Fig.\ \ref{cpdw}(a). In the vicinity of hot spots Eq.\ (\ref{25})  becomes
\begin{align}
S_{\rm int} =  i\sigma^y_{\alpha\beta}\varphi_A c_{1\alpha}^\dagger(\tilde k) c_{2\beta}^\dagger(-\tilde k)+i\sigma^y_{\alpha\beta}\varphi_{-A}c_{3\alpha}^\dagger(\tilde k)c_{4\beta}^\dagger(-\tilde k)+\delta_{\alpha\beta}\rho_A c^\dagger_{2\alpha}(\tilde k) c_{1\beta}(\tilde k)+\delta_{\alpha\beta} \rho_{-A}c^\dagger_{4\alpha}(\tilde k) c_{3\beta}(\tilde k)+h.c.
  \label{1}
  \end{align}
  where, we remind, $\tilde k$ is the deviation from a corresponding hot spot [not to be confused with $k$ in Eq.\ (\ref{25})], and we have defined at hot spots $\delta_{\alpha\beta}\rho_A\propto \langle c_{1\alpha}^\dagger c_{2\beta}\rangle$, $\delta_{\alpha\beta}\rho_{-A} \propto\langle c_{3\alpha}^\dagger c_{4\beta}\rangle$, $i\sigma_{\alpha\beta}^y\varphi_A\propto\langle c_{1\alpha}c_{2\beta}\rangle$, and $i\sigma_{\alpha\beta}^y\varphi_{-A}\propto\langle c_{3\alpha} c_{4\beta}\rangle$ [see Fig.\ \ref{cpdw}(a)].

We present the ladder equations for $\varphi$ and $\rho$ diagrammatically in Fig.\ \ref{cpdw}(b,c). The spin fluctuation propagator relates $\varphi_A,\rho_A$ with $\varphi_{-A},\rho_{-A}$, and vise versa. As our goal in this Section is to obtain the instability, we consider these equations to first order in the condensates $\rho$ and $\varphi$, and neglect feedback from the condensates to the fermionic propagators.

The self-consistent ladder equations for PDW and CDW orders are
  \begin{align}
  \varphi_{A}(\omega_m',\tilde k')=&-3\bar gT\sum_{m}\int\frac{d\tilde k^2}{4\pi^2}\chi(\omega_m'-\omega_m, \tilde k-\tilde k')G_3(\omega_m,\tilde k)G_4(-\omega_m,-\tilde k)\varphi_{-A}(\omega_m,\tilde k)\nonumber\\
    \varphi_{-A}(\omega_m',\tilde k')=&-3\bar gT\sum_{m}\int\frac{d\tilde k^2}{4\pi^2}\chi(\omega_m'-\omega_m, \tilde k-\tilde k')G_1(\omega_m,\tilde k)G_2(-\omega_m,-\tilde k)\varphi_{A}(\omega_m,\tilde k),
  \label{3}
  \end{align}
  and
    \begin{align}
  \rho_{A}(\omega_m',\tilde k')=&3\bar gT\sum_{m}\int\frac{d\tilde k^2}{4\pi^2}\chi(\omega_m'-\omega_m, \tilde k-\tilde k')G_3(\omega_m,\tilde k)G_4(\omega_m,\tilde k)\rho_{-A}(\omega_m,\tilde k), \nonumber\\
 \rho_{-A}(\omega_m',\tilde k')=&3\bar gT\sum_{m}\int\frac{d\tilde k^2}{4\pi^2}\chi(\omega_m'-\omega_m, \tilde k-\tilde k')G_1(\omega_m,\tilde k)G_2(\omega_m,\tilde k)\rho_{A}(\omega_m,\tilde k)
  \label{4}
  \end{align}
  where $G_i(\omega_m,\tilde k)=-1/[i\omega_m+\Sigma_i(\omega_m)-\epsilon_{i,\tilde k}]$ is the fermionic Green's function and we defined
 \begin{align}
\chi(q)=\frac{1}{{\bf q}^2+\xi^{-2}+\Pi(\Omega_m)}=\frac{1}{{\bf q}^2+\xi^{-2}+\gamma|\Omega_m|}.
\label{chi}
\end{align}
  The prefactors $-3$ and $3$ come from summing over spin indices for the PDW channel and for the CDW channel (for PDW $i\sigma^y_{\alpha\beta}\vec\sigma_{\alpha\gamma}\cdot\vec\sigma_{\beta\delta}=-3i\sigma^{y}_{\gamma\delta}$, while for CDW  $\delta_{\alpha\beta}\vec\sigma_{\alpha\gamma}\cdot\vec\sigma_{\delta\beta}=3\delta_{\gamma\delta}$).
   For the linearized fermionic dispersion, $\epsilon_{i,\tilde k}$ is odd in momentum deviation $\tilde k$ from hot spot, and we have $G_i(\omega_m,\tilde k)=-G_i(-\omega_m,-\tilde k)$.
   Substituting this into Eq.\ (\ref{3}) and comparing with Eq.\ (\ref{4}) we find that
    the minus sign from changing the signs of $\omega$ and ${\tilde k}$ in one of the Green's function in (\ref{3}) compensates the difference in the
       overall factors due to spin summation and, as a result,  Eqs.\ (\ref{3}) and (\ref{4}) become exactly the same. Eq.\ (\ref{4}) has been studied in detail
 (see Sec. III in Ref.\ \onlinecite{charge}) and was shown to give rise to a CDW instability at a nonzero temperature $T_{\rm CDW}$. By the same reasoning, Eq.\ (\ref{3}) should yield an instability towards PDW order at the same temperature $T_{\rm PDW}=T_{\rm CDW}$.

 That $T_{\rm PDW}=T_{\rm CDW}$ is non-zero can be understood from the following scaling arguments. At the magnetic critical point
 the bosonic propagator scales as $\chi \propto \chi_0/(q^2+ \gamma|\omega|)$ and the fermionic propagator scales as $G \propto 1/(i\sqrt {\omega_m \omega_0} -v_F q)$ (neglecting numerical prefactors). Rescaling $\omega$ by $\omega_0 \propto {\bar g}$ and ${\bf q}$ by ${\bar g}/v_F$ we find after simple algebra
 that all dimensional factors
    in the r.h.s.\ of Eqs.\ (\ref{3}) and (\ref{4}) cancel out and   ${\bar g} \int \chi GG$ scales as
    $\int d^2xdy 1/[(i\sqrt {y} -x)^2 (x^2 + |y|)] \sim \int dz dy/(z+y)^2$. This integral diverges logarithmically.
       Taking lower limit of the frequency integration as $T$ and the upper as ${\bar g}$ (at higher frequencies self-energy is irrelevant)
        we find that the r.h.s.\ of Eqs.\ (\ref{3}) and (\ref{4}) scales as $\log(\bar g/T)$ (for the details on evaluation of the integrals see Appendix B of
         Ref. \onlinecite{charge}).  We then obtain for either CDW or PDW order
         \begin{align}
\varphi_A=- S_1 \log{\frac{\bar g}{T}} \varphi_{-A},&~~~  \varphi_{-A}=- S_2 \log{\frac{\bar g}{T}} \varphi_{A},\nonumber\\
\rho_A=- S_1 \log{\frac{\bar g}{T}} \rho_{-A},&~~~  \rho_{-A}=- S_2 \log{\frac{\bar g}{T}} \rho_{A}.
\label{7_1}
\end{align}
where $S_1>0$ and $S_2>0$ are numerical prefactors which depend on the ratio of $v_x/v_y$
(for $v_x =0$,  $S_1=0.084$ and $S_2=0.650$).  Because of logarithms, the set (\ref{7_1}) has a non-trivial solution
 at
 \begin{align}
 T_{\rm PDW}=T_{\rm CDW} \sim {\bar g}  e^{-1/\sqrt{S_1 S_2}}.
 \label{ac_last}
 \end{align}
  We also see from (\ref{7_1}) that $\varphi_A$ and $\varphi_{-A}$  and $\rho_A$ and $\rho_{-A}$  should have opposite signs due to the repulsive nature of the spin-fermion interaction:
\begin{align}
\varphi_{-A}=&-\lambda\varphi_{A}\nonumber\\
\rho_{-A}=&-\lambda\rho_{A}.
\label{7}
\end{align}
where $\lambda = \sqrt{S_2/S_1}$. Evaluating $S_1$ and $S_2$ one finds that  $S_2 \geq S_1$, hence $\lambda > 1$ (see Sec.\ III of Ref.\ \onlinecite{charge}). We recall that the hot regions with $\varphi_A,\rho_A$ and with $\varphi_{-A},\rho_{-A}$ differ in momentum by $(\pi,\pi)$. Eq.\ (\ref{7}) then implies that both CDW and PDW orders have a form factor
 which changes sign under the shift by $(\pi,\pi)$. At the same time, the magnitudes of $\varphi_A$ and $\varphi_{-A}$ and of $\rho_A$ and $\rho_{-A}$ are not equal,
   unless $S_2 = S_1$.  The implication is that the form factor for CDW and PDW has both  a $d$-wave component and an $s$-wave component, and the restriction set by Eq.\ (\ref{7}) is that $d$-wave component is larger~\cite{laplaca,charge,davis_1}. A pure $d$-wave form-factor is recovered in the limit  $S_1 =S_2$.
    For simplicity, below we will be referring to CDW and PDW form-factors as  ``$d$-wave" just to emphasize that
    the  order parameters at $A$ and $-A$ must have opposite sign.

The doping dependence of $T_{\rm CDW}=T_{\rm PDW}\equiv T(x)$ can be studied within our model by varying the correlation length $\xi$
 and the chemical potential $\mu$. By varying the chemical potential $\mu$, one varies the position of hot spots, the CDW wave vector (i.e. the distance between hot spots), and the ratio $v_x/v_y$. Because  $S_1$ and $S_2$ are functions of $v_x/v_y$, $T_{\rm CDW}=T_{\rm PDW}\sim {\bar g}  e^{-1/\sqrt{S_1 S_2}}$ is generally affected. However, we found that $v_x/v_y$  depends on $\mu$ only weakly and hence the variation of $T_c$ is quite small (see Appendix \ref{ratio} for details).
 The variation with $\xi$ is far stronger as at finite $\xi$ the logarithm is cut-off at small $T$ and becomes $\log[\bar g/(T+\xi^{-2}/\gamma)]$.
  As a result, $T_{\rm PDW}$ and $T_{\rm CDW}$ decrease with increasing $\xi^{-1}$ and vanish at some critical $\xi^{-1}_{cr} \sim\bar g/v_F$.
 We use this fact when we construct the phase diagram.

\subsection{PDW and CDW as intertwined orders from ${\rm SU}(2)$ particle-hole symmetry}
MS pointed out that there exists a hidden ${\rm SU}(2)$ particle-hole symmetry in the spin-fermion model with linear fermionic dispersion~\cite{ms}. Below we reproduce their result using slightly different notations and then use this symmetry to reveal the degeneracy between CDW and PDW orders.

First we introduce eight ``pseudo-spinors", each at a given hot spot,
\begin{align}
&\Psi_1({\tilde k})=\(\begin{array}{c}
c_{1\uparrow}({\tilde k})\\
c_{1\downarrow}^{\dagger}(-{\tilde k})
\end{array}\),~~
\Psi_2({\tilde k})=\(\begin{array}{c}
c_{2\downarrow}^\dagger(-{\tilde k})\\
c_{2\uparrow}({\tilde k})
\end{array}\)\nonumber\\
&\Psi_3({\tilde k})=\(\begin{array}{c}
c_{3\uparrow}({\tilde k})\\
c_{3\downarrow}^{\dagger}(-{\tilde k})
\end{array}\),~~
\Psi_4({\tilde k})=\(\begin{array}{c}
c_{4\downarrow}^\dagger(-{\tilde k})\\
c_{4\uparrow}({\tilde k})
\end{array}\)\nonumber\\
&\Psi_5({\tilde k})=\(\begin{array}{c}
c_{5\downarrow}^\dagger(-{\tilde k})\\
c_{5\uparrow}({\tilde k})
\end{array}\),~~
\Psi_6({\tilde k})=\(\begin{array}{c}
c_{6\uparrow}({\tilde k})\\
c_{6\downarrow}^{\dagger}(-{\tilde k})
\end{array}\)\nonumber\\
&\Psi_7({\tilde k})=\(\begin{array}{c}
c_{7\downarrow}^\dagger(-{\tilde k})\\
c_{7\uparrow}({\tilde k})
\end{array}\),~~
\Psi_8({\tilde k})=\(\begin{array}{c}
c_{8\uparrow}({\tilde k})\\
c_{8\downarrow}^{\dagger}(-{\tilde k})
\end{array}\),
\end{align}
where ${\tilde k}\equiv(\omega,{\bf {\tilde k}})$ and ${\tilde k}$ is momentum  deviation from the corresponding hot spot.
In this notation, the fermionic part of the action can be rewritten as
\begin{align}
\mathcal{S}_0=\sum_{ij,\mu\nu}\int d{\tilde k} \Psi_{i\mu}^\dagger(\tilde k)(-i\omega+\epsilon_{i,\tilde k})\delta_{ij}\delta_{\mu\nu}\Psi_{j\nu}(\tilde k),
\label{s0}
\end{align}
where $ij$ label hot spots,  $\mu\nu$ are pseudo-spin indices, and we remind that $\epsilon_{i,\tilde k}$ is the linearized fermionic dispersion.
The pseudo-spin ${\rm SU}(2)$ symmetry is explicit in $\mathcal{S}_0$.  To see the ${\rm SU}(2)$ symmetry for the full action, we rewrite the fermionic fields $c$'s in Eq.\ (\ref{sf}) in terms of $\Psi$'s and obtain,
\begin{align}
\mathcal{S}=&\sum_{ij,\mu\nu}\int d\tilde k \Psi_{i\mu}^\dagger(\tilde k)(-i\omega+\epsilon_{i,\tilde k})\delta_{ij}\delta_{\mu\nu}\Psi_{j\nu}(\tilde k)+\frac{1}{2}\int dq\chi_0^{-1}(q)\vec\phi(q)\cdot\vec\phi(-q)\nonumber\\
&+g\sum_{\substack{i=1,2,5,6\\\mu\nu}}\int d\tilde kd\tilde k'\[\Psi_{i\mu}^\dagger(\tilde k)\delta_{\mu\nu}\Psi_{i+2,\nu}(\tilde k')+\Psi_{i+2,\mu}^\dagger(\tilde k)\delta_{\mu\nu}\Psi_{i,\nu}(\tilde k')\]\phi_z(\tilde k-\tilde k')\nonumber\\
&+g\sum_{\substack{i=1,2,5,6\\\mu\nu}}\int d\tilde kd\tilde k'\[\Psi_{i\mu}^\dagger(\tilde k)(-i\sigma_{\mu\nu}^y)\Psi_{i+2,\nu}^+(-\tilde k')+\Psi_{i+2,\mu}^\dagger(\tilde k)(-i\sigma_{\mu\nu}^y)\Psi_{i,\nu}^+(-\tilde k')\]\phi_x(\tilde k-\tilde k')\nonumber\\
&+g\sum_{\substack{i=1,2,5,6\\\mu\nu}}\int d\tilde kd\tilde k'\[\Psi_{i\mu}^\dagger(\tilde k)(\sigma_{\mu\nu}^y)\Psi_{i+2,\nu}^+(-\tilde k')+\Psi_{i+2,\mu}^\dagger(\tilde k)(\sigma_{\mu\nu}^y)\Psi_{i,\nu}^+(-\tilde k')\]\phi_y(\tilde k-\tilde k'),
\label{su2}
\end{align}
where $\tilde k,\tilde k'$ are deviations from corresponding hot spots and the $\Psi^+$ in last two lines denotes taking the Hermitian conjugate in the Fock space without transposing in pseudo-spin space,
e.g.,
\begin{align}
\Psi^+_1 (k) = \(\begin{array}{c}
c^\dagger_{1\uparrow}({\tilde k})\\
c_{1\downarrow}(-{\tilde k})
\end{array}\).
\end{align}
Eq.\ (\ref{su2}) is now explicitly invariant under four independent ${\rm SU}(2)$ pseudo-spin rotations.
\begin{align}
\Psi_{i}\rightarrow \mathcal{U}_i\Psi_{i},~\Psi_{i+2}\rightarrow \mathcal{U}_i \Psi_{i+2},
\label{4su2}
\end{align}
where $i=1,2,5,6$ and $\mathcal{U}_i$'s are generic ${\rm SU}(2)$ matrices. To see the invariance it is helpful to use the relations $\Psi_i^+\rightarrow \mathcal{U}_i^* \Psi^+$, and $\mathcal{U}_i^\dagger\sigma^y\mathcal{U}_i^*=\sigma^y$.

We can now rewrite Eq.\ (\ref{1}) as
\begin{align}
S^A_{\rm int} = \Psi^\dagger_{1\mu}(\tilde k)\Delta^{\mu\nu}_{A}\Psi_{2\nu}(\tilde k)+h.c.,
\label{thur}
\end{align}
and define a PDW/CDW condensate $\Delta_A$ that couples bilinearly to the pseudo-spinor fields $\Psi_1$ and $\Psi_2$:
\begin{align}
\Delta_A^{\mu\nu}&=\(\begin{array}{cc}\varphi_A&\rho_A^*\\-\rho_A&\varphi_A^*\end{array}\)
=\sqrt{|\rho_A|^2+|\varphi_A|^2}\(\begin{array}{cc}
\varphi_A^*/\sqrt{|\rho_A|^2+|\varphi_A|^2}&\rho_A^*/\sqrt{|\rho_A|^2+|\varphi_A|^2}\\
-\rho_A/\sqrt{|\rho_A|^2+|\varphi_A|^2}&\varphi_A/\sqrt{|\rho_A|^2+|\varphi_A|^2}
\end{array}\)\nonumber\\
&\equiv\sqrt{|\rho_A|^2+|\varphi_A|^2}~U_A,
\label{144}
\end{align}
  where $U_A$ is an ${\rm SU}(2)$ ``phase". When $U_A$ is diagonal, then the system has PDW order and when it is anti-diagonal, the system has CDW order. Under an ${\rm SU}(2)$ pseudo-spin rotation, the CDW and PDW mix with each other.  A self consistent equation for a CDW/PDW condensate with a generic ${\rm SU}(2)$ ``phase" can be straightforwardly derived directly from Eq.\ (\ref{su2}), in terms of pseudo-spinor $\Psi$'s. As expected, it coincides with Eqs.\ (\ref{3}) and (\ref{4}), which, we remind, are identical.
We will keep the symmetry between CDW and PDW explicit in the next Section and describe both order parameters using a combined  PDW/CDW order parameter $\Delta$.

\section{Effective action for the PDW/CDW order parameter}
In this Section we derive the effective action for PDW and CDW order parameters in an ${\rm SU}(2)$-covariant form. We apply a Hubbard-Stratonovich transformation to the spin-fermion model  to decouple the effective four-fermion interactions $\sim \Psi^\dagger\Psi\Psi^\dagger\Psi$ into bilinear couplings between a new bosonic field and fermions $\sim \Psi^\dagger \Delta \Psi$. We then integrate out the fermion field $\Psi$'s to obtain the effective Ginzburg-Landau (GL) action in terms of $\Delta$'s
  which in this Section we treat as  fluctuating fields rather than condensates.
  At low temperatures, the minimization of the GL action yields nonzero
   condensate
   values for $\Delta$'s and the system develops a PDW/CDW order.
   The condensate values obtained this way are equivalent to the ones which one would obtain by solving non-linear ladder equations for $\varphi$ and $\rho$
   (same as in the previous Section but extended  to finite $\varphi$ and $\rho$.)

\begin{figure}
\includegraphics[width=0.4\columnwidth]{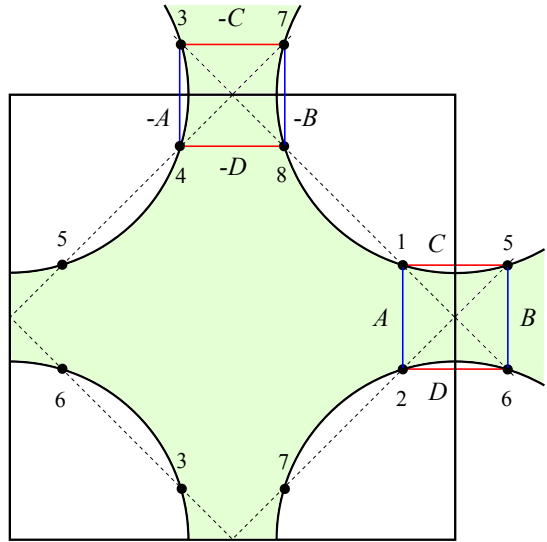}
\caption{the Brillouin zone, the magnetic Brillouin zone (dashed line), and the Fermi surface of a cuprate. The hot spots 1-8 are defined as the interSection of hot spots and the Brillouin zone. We label ``bonds" connecting hot spots by $A,B,C,D$ and $-A,-B,-C,-D$. Bonds connecting hot spots 1256 and 3478 are denoted by opposite signs, due to the $d$-form factor.}
\label{bz}
\end{figure}

We label ``bonds" connecting hot spots 1-8 by $A,B,C,D$ and corresponding bonds with momenta shifted by $(\pi,\pi)$ by $-A,-B,-C,-D$ (see  Fig.\ \ref{bz}).
Through a Hubbard-Stratonovich transformation we introduce four PDW/CDW order parameters $\Delta_{A,B,C,D}$ which couple bilinearly to fermions as
$S^A_{\rm int}=\Psi^\dagger_{1\mu}(\tilde k)\Delta^{\mu\nu}_{A}\Psi_{2\nu}(\tilde k)$, $S^B_{\rm int} = \Psi^\dagger_{6\mu}(\tilde k)\Delta^{\mu\nu}_{B}\Psi_{5\nu}(\tilde k)$, $S^C_{\rm int} = -\Psi^\dagger_{3\mu}(\tilde k)\Delta^{\mu\nu}_{C}\Psi_{7\nu}(\tilde k)$, and $S^D_{\rm int} = -\Psi^\dagger_{8\mu}(\tilde k)\Delta^{\mu\nu}_{D}\Psi_{4\nu}(\tilde k)$. Similar to Eq.\ (\ref{144}), each PDW/CDW order parameter has PDW and CDW components:
\begin{align}
\Delta_A^{\mu\nu}=\(\begin{array}{cc}
\varphi_A&\rho_A^*\\
-\rho_A&\varphi_A^*
\end{array}\),~
\Delta_B^{\mu\nu}=\(\begin{array}{cc}
\varphi_B&\rho_B\\
-\rho_B^*&\varphi_B^*
\end{array}\),~
\Delta_C^{\mu\nu}=\(\begin{array}{cc}
\varphi_C&\rho_C^*\\
-\rho_C&\varphi_C^*
\end{array}\),~\Delta_D^{\mu\nu}=\(\begin{array}{cc}
\varphi_D&\rho_D\\
-\rho_D^*&\varphi^*_D
\end{array}\),
\label{1106}
\end{align}
where, for example $\rho_A\sim c_1^\dagger c_2$, $\varphi_A\sim c_1c_2$, $\rho_B\sim c_5^\dagger c_6$, and $\varphi_B\sim c_5c_6$.
 It is easy to verify that under time reversal, $\langle\rho_{A,C}\rangle\rightarrow\langle\rho_{B,D}\rangle$ and $\langle\varphi_{A,C}\rangle\rightarrow\langle\varphi_{B,D}^*\rangle$, therefore, under time reversal, $\langle\Delta_{A,C}\rangle\rightarrow\langle\Delta_{B,D}^*\rangle$.

We remind that the bonds denoted by the same letter (e.g. $A$ and $-A$) have order parameters of opposite sign, and differ in magnitude by a factor of $\lambda$. Using this relation and lattice symmetries
 we can
  write the effective  action for fermions and  PDW/CDW order parameters
  in a covariant form as
\begin{align}
\mathcal{S}=&\Psi_{i\mu}^\dagger(-i\omega+\epsilon_{i})\delta_{ij}\delta_{\mu\nu}\Psi_{j\nu}+\alpha_0 \Tr(\Delta_A^\dagger \Delta_A+\Delta_B^\dagger \Delta_B+\Delta_C^\dagger \Delta_C+\Delta_D^\dagger \Delta_D)\nonumber\\
&+\Psi_{1\mu}^\dagger\Delta_A^{\mu\nu}\Psi_{2\nu}+\Psi_{6\mu}^\dagger\Delta_B^{\mu\nu}\Psi_{5\nu}+\lambda\Psi_{1\mu}^\dagger\Delta_C^{\mu\nu}\Psi_{5\nu}+\lambda\Psi_{6\mu}^\dagger\Delta_D^{\mu\nu}\Psi_{2\nu}\nonumber\\
&-\lambda\Psi_{3\mu}^\dagger\Delta_A^{\mu\nu}\Psi_{4\nu}-\lambda\Psi_{8\mu}^\dagger\Delta_B^{\mu\nu}\Psi_{7\nu}-\Psi_{3\mu}^\dagger\Delta_C^{\mu\nu}\Psi_{7\nu}-\Psi_{8\mu}^\dagger\Delta_D^{\mu\nu}\Psi_{4\nu}+h.c.
\label{sd}
\end{align}
where for
compactness
we have omitted the symbols of
 momentum and frequency integrations, which are assumed in (\ref{sd}).
Eq.\ (\ref{sd}) can be derived directly from the spin-fermion model by first integrating out $\phi$ fields to get effective four-fermion interaction and then applying a Hubbard-Stratonovich transformation to decouple the interaction (see Sec.\ IV\ B and Appendix D of Ref. \onlinecite{charge}).

Since $\Psi_\mu$'s transform as ${\rm SU}(2)$, then $\Delta^{\mu\nu}$'s, each of which couples to $\Psi$'s through $\Psi_{\mu}^\dagger \Delta^{\mu\nu}\Psi_{\nu}$, must transform as ${\rm SU}(2)\times {\rm SU}(2)$, which is
homomorphic to ${\rm SO}(4)$ (see e.g.\ Ref. \onlinecite{fulton}). We
 will  see this ${\rm SO}(4)$ symmetry explicitly below.

 Because Eq.\ (\ref{sd}) is bilinear in fermionic operators,  one can explicitly integrate out
  fermions and obtain the effective action in terms of $\Delta$'s.
For small $\Delta$, one can expand the effective action perturbatively in powers of $\Delta$. First, at order $\Delta^2$, fermionic bubbles formed between hot spots $i$ and $i+2$ $(i=1,2,5,6)$ renormalize the coefficient $\alpha_0$ in (\ref{sd}) to $\alpha(T)=\alpha_0-A(T)$, where $A(T)$ increases upon lowering temperature. At $T=T_{\rm CDW}=T_{\rm PDW}$, $\alpha$ becomes zero and the system develops an instability towards CDW/PDW order.

\begin{figure*}
\includegraphics[width=0.7\columnwidth]{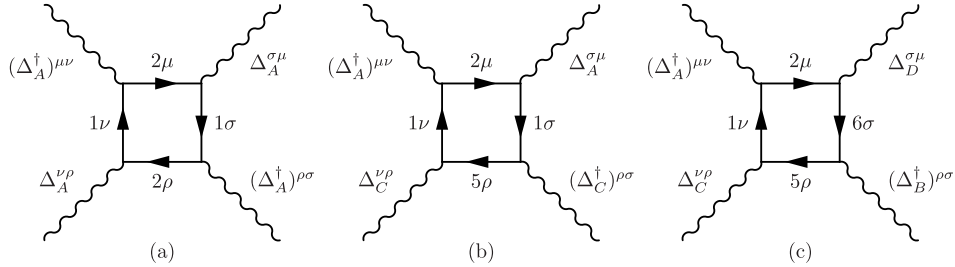}
\caption{Diagrammatic representation for fourth order terms of the CDW/PDW order parameter.}
\label{2}
\end{figure*}

The contribution to the effective action at order $\Delta^4$ comes from square diagrams shown in Fig.\ \ref{2}.
The Green's function for $\Psi$ can be straightforwardly obtained from Eq.\ (\ref{s0}):
\begin{align}
\langle T\Psi_{i\mu}\Psi_{j\nu}^\dagger \rangle = {\delta_{\mu\nu}\delta_{ij}}G_i(\tilde k) = -\frac{\delta_{\mu\nu}\delta_{ij}}{i\omega-\epsilon_i(\tilde k)}.
\end{align}
Then the diagram in Fig.\ \ref{2}(a) is expressed as
\begin{align}
\mathcal{S}_{a}^A=&\frac{1}{2}\Tr(\Delta_A\Delta_A^\dagger\Delta_A\Delta_A^\dagger)\int d\tilde k G_1^2(\tilde k)G_2^2(\tilde k)\nonumber\\
=&(|\rho_A|^2+|\varphi_A|^2)^2\int d\tilde k G_1^2(\tilde k)G_2^2(\tilde k).
\end{align}
With $\rho_A$ and $\varphi_A$ each being a complex field, we see explicitly the ${\rm SO}(4)$   symmetry between CDW and PDW components. Summing contributions of this type from $\Delta_A$, $\Delta_B$, $\Delta_C$ and $\Delta_D$, we obtain
\begin{align}
\mathcal{S}_{a}=&-(I_1+\lambda^4I_2)\Tr(\Delta_A\Delta_A^\dagger\Delta_A\Delta_A^\dagger+\Delta_B\Delta_B^\dagger\Delta_B\Delta_B^\dagger+\Delta_C\Delta_C^\dagger\Delta_C\Delta_C^\dagger+\Delta_D\Delta_D^\dagger\Delta_D\Delta_D^\dagger)\nonumber\\
=&-2(I_1+\lambda^4I_2)\[(|\rho_A|^2+|\varphi_A|^2)^2+(|\rho_B|^2+|\varphi_B|^2)^2+(|\rho_C|^2+|\varphi_C|^2)^2+(|\rho_D|^2+|\varphi_D|^2)^2\]
\end{align}
where $I_1=-\frac{1}{2}\int d\tilde k G_1^2(\tilde k)G_2^2(\tilde k)<0$ and $I_2=-\frac{1}{2}\int d\tilde k G_3^2(\tilde k)G_4^2(\tilde k)<0$ (Ref.\ \onlinecite{charge}). By construction, the integrals $I_1$ and $I_2$ are confined to the vicinity of hot spots, therefore at this stage there is no coupling between order parameters at different hot spots, e.g., $\Delta_A$ and $\Delta_B$.

Evaluating the diagram in Fig.\ \ref{2}(b), we obtain
\begin{align}
\mathcal{S}_{b}^{AC}=&\lambda^2\Tr(\Delta_A\Delta_A^\dagger\Delta_C\Delta_C^\dagger)\int dk G_1^2(k)G_2(k)G_5(k)\nonumber\\
=&2\lambda^2(|\rho_A|^2+|\varphi_A|^2)(|\rho_C|^2+|\varphi_C|^2)\int dk G_1^2(k)G_2(k)G_5(k),
\end{align}
where again the ${\rm SO}(4)$  symmetry is explicit. Summing  contributions from same type of diagrams involving $(\Delta_A, \Delta_D)$,  $(\Delta_B, \Delta_C)$, and  $(\Delta_B, \Delta_D)$  we obtain
\begin{align}
\mathcal{S}_{b}=&-2\lambda^2I_3\Tr\[(\Delta_A\Delta_A^\dagger+\Delta_B\Delta_B^\dagger)(\Delta_C\Delta_C^\dagger+\Delta_D\Delta_D^\dagger)\]\nonumber\\
=&-\lambda^2I_3(|\rho_A|^2+|\varphi_A|^2+|\rho_B|^2+|\varphi_B|^2)(|\rho_C|^2+|\varphi_C|^2+|\rho_D|^2+|\varphi_D|^2),
\end{align}
where $I_3=-\int dk G_1^2(\tilde k)G_2(\tilde k)G_5(\tilde k)<0$ (Ref.\ \onlinecite{charge}).

Finally, the diagram in Fig.\ \ref{2}(c) together with its conjugate yields
\begin{align}
\mathcal{S}_{c}=&-2\lambda^2I_4\Tr\[\Delta_A^\dagger\Delta_B\Delta_C^\dagger\Delta_D\]+h.c.\nonumber\\
=&-2\lambda^2I_4\sqrt{(|\rho_A|^2+|\varphi_A|^2)(|\rho_B|^2+|\varphi_B|^2)(|\rho_C|^2+|\varphi_C|^2)(|\rho_D|^2+|\varphi_D|^2)}\[\Tr\(U_A^\dagger U_CU_B^\dagger U_D\)+h.c.\],
\label{11}
\end{align}
where $I_4=-\int d\tilde k G_1(\tilde k)G_2(\tilde k)G_5(\tilde k)G_6(\tilde k)<0$
 (Ref.\ \onlinecite{charge}), and in the last line we
defined ${\rm SU}(2)$ phases for $\Delta_{B,C,D}$ the same way as in Eq.\ (\ref{144}), namely,
\begin{align}
\Delta_{A,B,C,D}\equiv\sqrt{|\rho_{A,B,C,D}|^2+|\varphi_{A,B,C,D}|^2}~U_{A,B,C,D}.
\end{align}

Summing up all terms up to $O(\Delta^4)$, we obtain the Ginzburg-Landau action
\begin{align}
\mathcal{S}_{\rm eff}=&\alpha \Tr(\Delta_A^\dagger \Delta_A+\Delta_B^\dagger \Delta_B+\Delta_C^\dagger \Delta_C+\Delta_D^\dagger \Delta_D)+\mathcal{S}_{a}+\mathcal{S}_{b}+\mathcal{S}_{c}\nonumber\\
=&\alpha \Tr(\Delta_A^\dagger \Delta_A+\Delta_B^\dagger \Delta_B+\Delta_C^\dagger \Delta_C+\Delta_D^\dagger \Delta_D)\nonumber\\
&-(I_1+\lambda^4I_2)\Tr(\Delta_A\Delta_A^\dagger\Delta_A\Delta_A^\dagger+\Delta_B\Delta_B^\dagger\Delta_B\Delta_B^\dagger+\Delta_C\Delta_C^\dagger\Delta_C\Delta_C^\dagger+\Delta_D\Delta_D^\dagger\Delta_D\Delta_D^\dagger)\nonumber\\
&-2\lambda^2I_3\Tr\[(\Delta_A\Delta_A^\dagger+\Delta_B\Delta_B^\dagger)(\Delta_C\Delta_C^\dagger+\Delta_D\Delta_D^\dagger)\]-2\lambda^2I_4\Tr\[\Delta_A^\dagger\Delta_B\Delta_C^\dagger\Delta_D\]+h.c.\nonumber\\
=&2\alpha (|\rho_A|^2+|\varphi_A|^2+|\rho_B|^2+|\varphi_B|^2+|\rho_C|^2+|\varphi_C|^2+|\rho_D|^2+|\varphi_D|^2)\nonumber\\
&-2(I_1+\lambda^4I_2)\[(|\rho_A|^2+|\varphi_A|^2)^2+(|\rho_B|^2+|\varphi_B|^2)^2+(|\rho_C|^2+|\varphi_C|^2)^2+(|\rho_D|^2+|\varphi_D|^2)^2\]\nonumber\\
&-4\lambda^2I_3(|\rho_A|^2+|\varphi_A|^2+|\rho_B|^2+|\varphi_B|^2)(|\rho_C|^2+|\varphi_C|^2+|\rho_D|^2+|\varphi_D|^2)\nonumber\\
&-2\lambda^2I_4\sqrt{(|\rho_A|^2+|\varphi_A|^2)(|\rho_B|^2+|\varphi_B|^2)(|\rho_C|^2+|\varphi_C|^2)(|\rho_D|^2+|\varphi_D|^2)}\[\Tr\(U_A^\dagger U_CU_B^\dagger U_D\)+h.c.\]
\label{seff}
\end{align}

 Because the spin-fermion model has four independent ${\rm SU}(2)$ symmetries [see Eq.\ (\ref{4su2})], the full effective action (\ref{seff}) should have an ${\rm SU}(2)\times{\rm SU}(2)\times{\rm SU}(2)\times{\rm SU}(2)\sim{\rm SO}(4)\times {\rm SO}(4)$ symmetry. We
  present the mathematical proof
   of ${\rm SO}(4)\times {\rm SO}(4)$ symmetry
  in Appendix \ref{app:1}.

\section{The structure of the ground state configuration \label{Sec4}}
In this Section we minimize Eq.\ (\ref{seff}) with respect to four order parameters $\Delta_{A,B,C,D}$ and obtain the condensate values of $\varphi$ and $\rho$.
We first rewrite Eq.\ (\ref{seff}) as
\begin{align}
\mathcal{S}_{\rm eff}=&2\alpha (|\rho_A|^2+|\varphi_A|^2+|\rho_B|^2+|\varphi_B|^2+|\rho_C|^2+|\varphi_C|^2+|\rho_D|^2+|\varphi_D|^2)\nonumber\\
&+\frac{1}{2}\beta\[(|\rho_A|^2+|\varphi_A|^2)^2+(|\rho_B|^2+|\varphi_B|^2)^2+(|\rho_C|^2+|\varphi_C|^2)^2+(|\rho_D|^2+|\varphi_D|^2)^2\]\nonumber\\
&+\frac{1}{4}\tilde\beta_m(|\rho_A|^2+|\varphi_A|^2+|\rho_B|^2+|\varphi_B|^2)(|\rho_C|^2+|\varphi_C|^2+|\rho_D|^2+|\varphi_D|^2)\nonumber\\
&+\frac{1}{2}\bar\beta_m\gamma\sqrt{(|\rho_A|^2+|\varphi_A|^2)(|\rho_B|^2+|\varphi_B|^2)(|\rho_C|^2+|\varphi_C|^2)(|\rho_D|^2+|\varphi_D|^2)}.
\label{13}
\end{align}
where $\beta=-4(I_1+\lambda^4I_2)>0$, $\tilde\beta_m=-16\lambda^2I_3>0$, and $\bar\beta_m=-8\lambda^2I_4>0$, and $\gamma=\Tr\(U_A^\dagger U_CU_B^\dagger U_D\)$. The matrix product $U_A^\dagger U_CU_B^\dagger U_D$ is still an ${\rm SU}(2)$  matrix, and $\gamma$  satisfies $-2\leq\gamma\leq2$. Minimization with respect to  the last line in (\ref{13}) then yields $\gamma=-2$.
 When written in terms of
  $\varphi$ and $\rho$, using
 \begin{align}
 U_i=\frac{1}{\sqrt{|\rho_i|^2+|\varphi_i|^2}}\(\begin{array}{cc}
\varphi_i&\rho_i^*\\
-\rho_i&\varphi_i^*
\end{array}\),
\end{align}
 the condition $\gamma=-2$ becomes
 \begin{align}
\frac{\left(\varphi_A^*\varphi_C+\rho_A^*\rho_C\right)\left(\varphi_B^*\varphi_D+\rho_B\rho_D^*\right)-\left(\varphi_A^*\rho_C^*-\rho_A^*\varphi_C^*\right)
\left(\varphi_B\rho_D^*-\rho_B^*\varphi_D\right)+h.c.}{\sqrt{{(|\varphi_A|^2+|\rho_A|^2)}{(|\varphi_C|^2+|\rho_C|^2)}{(|\varphi_B|^2+|\rho_B|^2)}{(|\varphi_D|^2+|\rho_D|^2)}}}=-2.
\label{mon}
\end{align}
 Using the fact that Eq.\ (\ref{13}) is symmetric with respect to $A\rightarrow B$ and $C\rightarrow D$ and that there is no repulsion between $A$ and $B$, and between $C$ and $D$, we immediately obtain that $|\rho_A|^2+|\varphi_A|^2=|\rho_B|^2+|\varphi_B|^2$ and $|\rho_C|^2+|\varphi_C|^2=|\rho_D|^2+|\varphi_D|^2$, and hence,
\begin{align}
\mathcal{S}_{\rm eff}=&4\alpha (|\rho_A|^2+|\varphi_A|^2+|\rho_C|^2+|\varphi_C|^2)\nonumber\\
&+\beta\[(|\rho_A|^2+|\varphi_A|^2)^2+(|\rho_C|^2+|\varphi_C|^2)^2\]\nonumber\\
&+(\tilde\beta_m-\bar\beta_m)(|\rho_A|^2+|\varphi_A|^2)(|\rho_C|^2+|\varphi_C|^2).
\label{13a}
\end{align}
The evaluation of the integrals $I_i$ (Refs.\ \onlinecite{charge,23}) yields $\beta\ll\bar\beta_m,\tilde\beta_m$. The ground state configuration then depends on the interplay between $\tilde\beta_m$ and $\bar\beta_m$.
The two are comparable in magnitude at the onset temperature of CDW/PDW order.
 In this Section we keep $\bar\beta_m,\tilde\beta_m\gg\beta$ and treat $\tilde\beta_m$ and $\bar\beta_m$ as the two input parameters. In the parameter space of $\bar\beta_m$ and $\tilde\beta_m$ we find two types of ground states.

If $\tilde\beta_m>\bar\beta_m$, Eq.\ (\ref{13}) is minimized if either $|\rho_A|^2+|\varphi_A|^2$ or $|\rho_C|^2+|\varphi_C|^2$ is set to zero. This breaks lattice rotational symmetry $C_4$ down to $C_2$ (see Fig.\ \ref{bz}). We label this state as state I. Borrowing jargon from a pure CDW state, we call this state a ``stripe" state. However, we remind that momenta carried by CDW and PDW order parameters on the same bond (i.e., between same hot spots) are {\it orthogonal} -- if CDW has $Q_x$, then PDW has $Q_y$. In this state, since one of $|\rho_A|^2+|\varphi_A|^2$ or $|\rho_C|^2+|\varphi_C|^2$ is zero, then the last line of Eq.\ (\ref{seff}) is zero, and the condition $\gamma=-2$ is relaxed.

If $\bar\beta_m>\tilde\beta_m$, Eq.\ (\ref{13}) is minimized if $|\rho_A|^2+|\varphi_A|^2=|\rho_C|^2+|\varphi_C|^2$. Since we have set $|\rho_A|^2+|\varphi_A|^2=|\rho_B|^2+|\varphi_B|^2$ and $|\rho_C|^2+|\varphi_C|^2=|\rho_D|^2+|\varphi_D|^2$, then for this state, $|\rho_A|^2+|\varphi_A|^2=|\rho_B|^2+|\varphi_B|^2=|\rho_C|^2+|\varphi_C|^2=|\rho_D|^2+|\varphi_D|^2$.
 This gives rise to a ``checkerboard" order. We label this state as state II.

We remind that Eq.\ (\ref{13a}) only fixes the amplitudes of the PDW/CDW condensates. For state I, there is no constraint on the ${\rm SU}(2)$ ``phases" of $\Delta_A$ and $\Delta_B$, since the term proportional to $\gamma$ vanishes (see Eq.\ (\ref{13})),
and for state II, the only constraint on the ${\rm SU}(2)$ ``phases" is $\gamma=\Tr(U_A^\dagger U_CU_B^\dagger U_D)=-2$. Therefore each state has an infinite number of members.

 Recalling that under time reversal $\Delta_A\rightarrow \Delta_B^*$ and hence $U_A\rightarrow U_B^*$, any member of the degenerate states with $\Delta_A\neq \Delta_B^*$ naturally breaks time reversal symmetry. However, we note that at this stage, the breaking of time-reversal symmetry is part of the breaking of a continuous ${\rm SO}(4)\times {\rm SO}(4)$ symmetry (by selecting specific $U_A$ and $U_B$), rather than the breaking of an additional discrete ${\mathbb Z}_2$ symmetry.

 The coefficients $\tilde\beta_m$ and $\bar\beta_m$ have been evaluated within the spin-fermion model~\cite{charge}.  Both depend of $T/\Lambda$, where
  $\Lambda$ is the energy cutoff for the spin-fermion model (of order $E_F$). We are interested in $T \sim T_{\rm CDW}$, which are of order of spin-fermion coupling $g$.
  For such $T$, two couplings  behave as $\bar\beta_m\sim \Lambda/g$ and $\tilde\beta_m\sim \log \Lambda/g$.  In the strict theoretical low-energy limit,
  $g << \Lambda$, hence $\tilde\beta_m<\bar\beta_m$ and the system develops a checkerboard CDW/PDW order (state II).  However, in the
   cuprates $g\sim E_F \sim \Lambda$, hence both state I and state II can develop.  Below we treat $\bar\beta_m$ and $\tilde \beta_m$ as two parameters of comparable magnitude and analyze both  state I and state II.

  \section{Breaking the ${\rm SO}(4)$ symmetry of the effective action}
\label{sec:5}
As we pointed out in the Introduction, the particle-hole ${\rm SU}(2)$  symmetry is only approximate. It relies on two crucial assumptions, 1) that order parameters couple only to hot fermions, and 2) that in each hot region one can linearize the fermionic dispersion. Once we go beyond either of these assumptions, the ${\rm SU}(2)$ symmetry will be broken, hence the emergent ${\rm SO}(4)\sim {\rm SU}(2)\times ${\rm SU}(2)$ $ symmetry of the PDW/CDW order will be broken also. In this Section we study separately the effects of going beyond linear fermionic dispersion and of going beyond hot spot approximation. We find that the curvature of the Fermi surface reduces $T_{\rm CDW}$ more than $T_{\rm PDW}$, favoring the PDW order. On the other hand, once we go beyond hot spot approximation, we find additional terms which couple the phases of CDW order parameters $\rho_A$ and $\rho_B$. These extra terms select CDW order which breaks a discrete ${\mathbb Z}_2$ time-reversal symmetry (i.e., $\rho_A=a+ib,~\rho_B=a-ib$, $b\neq0$).
This ${\mathbb Z}_2$ symmetry gets broken at a higher $T$ than what would be the onset temperature of CDW order and this symmetry breaking pushes $T_{\rm CDW}$ to higher value
compared to mean-field result. This effect tends to favor CDW order over PDW order. Which of the two effects wins depends on microscopic details of the dispersion and interactions away from hot spots.

 The analysis in this section is applicable to both stripe and checkerboard states.  In the next section we consider how one can additionally lower the energy of the checkerboard state due to  the fact that combined CDW/PDW order develops a secondary SC order.

\subsection{Going beyond linear dispersion}
\label{5A}
In this Section we study the effect of the Fermi surface curvature $\kappa$ on PDW and CDW transition temperatures.
To simplify calculations, we assume $\kappa\ll1$ and
consider the limiting case when Fermi velocities of fermions at e.g. points 1 and 2 (set $A$) are antiparallel while those at points 3 and 4 (set $-A$) are parallel.
A generic case
when Fermi velocities at 1,2 and 3,4 are neither parallel and antiparallel has been studied in Ref.\ \onlinecite{charge} and the
results are qualitatively similar to the limiting case we consider here.
We solve the linearized ladder equations for CDW and PDW condensates, Eqs.\ (\ref{3}) and (\ref{4}), in the presence of a small but finite $\kappa$.
 For fermionic dispersion in $G_i(\omega_m,\tilde k)=-1/[i\omega_m+\Sigma_i(\omega_m)-\epsilon_{i,\tilde k}]$, we use $\epsilon_{1,\tilde k}=v_F(\tilde k_y + \kappa \tilde k_x^2/k_F)$, $\epsilon_{2,\tilde k}=v_F(-\tilde k_y + \kappa \tilde k_x^2/k_F)$, and $\epsilon_{3,\tilde k}=\epsilon_{4,\tilde k}=-v_F(\tilde k_x + \kappa \tilde k_y^2/k_F)$. Plugging these into  Eqs.\ (\ref{3}) and (\ref{4}) and obtain integral equations for $\varphi_{\pm A}$ and $\rho_{\pm A}$ as functions of frequency and momentum.

We make one more simplification by
treating CDW and PDW order parameters as constants between hot spots (1, 2) and (3, 4), and we set external frequencies to zero and external  momenta
 to their values at hot spots (i.e., avoid solving integral equation in momentum and frequency).
 Again, previous study found~\cite{wang}  that this  does not affect the onset temperatures by more than a number.
 Using this simplification we obtain for the PDW order
\begin{align}
\varphi_{A}=\frac{-3\bar gT_{\rm PDW}}{4\pi^2}\sum_{m}\int&\frac{d\tilde k_x~d\tilde k_y}{\[i\tilde\Sigma(\omega_m)-v_y(\tilde k_x+\kappa \tilde k_y^2/k_F)\]\[-i\tilde\Sigma(\omega_m)+v_y(\tilde k_x-\kappa \tilde k_y^2/k_F)\]}\frac{\varphi_{-A}}{\tilde k_x^2+\tilde k_y^2+\gamma|\omega_m|}\label{tr5}\\
\varphi_{-A}=\frac{-3\bar gT_{\rm PDW}}{4\pi^2}\sum_{m}\int&\frac{d\tilde k_x~d\tilde k_y}{\[i\tilde\Sigma(\omega_m)-v_y(\tilde k_y-\kappa \tilde k_x^2/k_F)\]\[-i\tilde\Sigma(\omega_m)-v_y(\tilde k_y-\kappa \tilde k_x^2/k_F)\]}\frac{\varphi_{A}}{\tilde k_x^2+\tilde k_y^2+\gamma|\omega_m|}\label{tr6}
\end{align}
and for CDW order
\begin{align}
\rho_{A}=\frac{3\bar gT_{\rm CDW}}{4\pi^2}\sum_{m}\int&\frac{d\tilde k_x~d\tilde k_y}{\[i\tilde\Sigma(\omega_m)-v_y(\tilde k_x+\kappa \tilde k_y^2/k_F)\]\[i\tilde\Sigma(\omega_m)-v_y(\tilde k_x+\kappa \tilde k_y^2/k_F)\]}
\frac{\rho_{-A}}{\tilde k_x^2+\tilde k_y^2+\gamma|\omega_m|}\label{tr7}\\
\rho_{-A}=\frac{3\bar gT_{\rm CDW}}{4\pi^2}\sum_{m}\int&\frac{d\tilde k_x~d\tilde k_y}{\[i\tilde\Sigma(\omega_m)-v_y(\tilde k_y-\kappa\tilde  k_x^2/k_F)\]\[i\tilde\Sigma(\omega_m)+v_y(\tilde k_y+\kappa \tilde k_x^2/k_F)\]}
\frac{\rho_{A}}{\tilde k_x^2+\tilde k_y^2+\gamma|\omega_m|}.\label{tr8}
\end{align}
Assuming that $\kappa\ll1$, one can expand the integration kernel in $\kappa$ and then integrate over $\tilde k_x$.  By doing so we find,
\begin{align}
\varphi_{A}=&-S_1\log\frac{\omega_0}{T_{\rm PDW}}\varphi_{-A}+R_1\kappa^2\varphi_{-A},\nonumber\\
\varphi_{-A}=&-S_2\log\frac{\omega_0}{T_{\rm PDW}}\varphi_{A}-R_2\kappa^2\varphi_{A};\nonumber\\
\rho_{A}=&-S_1\log\frac{\omega_0}{T_{\rm CDW}}\rho_{-A}+3R_1\kappa^2\rho_{-A},\nonumber\\
\rho_{-A}=&-S_2\log\frac{\omega_0}{T_{\rm CDW}}\rho_{-A}+R_3\kappa^2\rho_{A},
\label{forty-one}
\end{align}
where
$ S_1=0.084$, $S_2=0.650$,
 and $R_1,R_2,R_3$ are all positive. The calculation is trivial, which we show in Appendix \ref{curv}.  To leading order in $\kappa$, we then have
\begin{align}
S_1S_2\log^2\frac{\omega_0}{T_{\rm PDW}}&=1+(-S_1R_2+S_2R_1)\kappa^2\nonumber\\
S_1S_2\log^2\frac{\omega_0}{T_{\rm CDW}}&=1+(S_1R_3+3S_2R_1)\kappa^2.
\end{align}
For positive $R_1,~R_2$ and $R_3$, one can immediately verify that $-S_1R_2+S_2R_1<S_1R_3+3S_2R_1$. Therefore, $T_{\rm PDW}>T_{\rm CDW}$.

\subsubsection{Properties of
 a pure PDW ground state}
\label{5B}
\begin{figure}
\includegraphics[width=0.4\columnwidth]{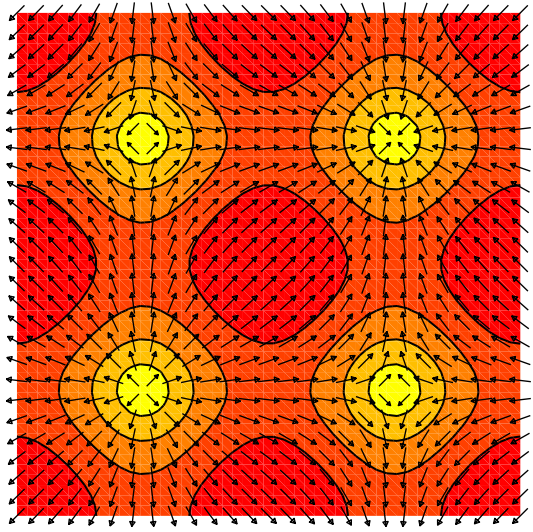}
\caption{The position dependence of the PDW representation of State II: the contour lines give the magnitude of the gap function while the $(x,y)$ components of the vectors denote the normalized real and imaginary parts of the gap function. Note the vortices and anti-vortices surrounding the zeroes of the gap function, giving a vortex-anti-vortex lattice.}
\label{PDW}
\end{figure}

Restricting the order parameter manifold to PDW states, together with the considerations of Sec.~\ref{Sec4} with respect to the parameters $\tilde\beta_m$ and $\bar\beta_m$, yields two possible ground states. The pure PDW representation of State I is a Larkin-Ovchinnikov (LO) state \cite{LO}, also known as a stripe SC state,  that has been studied by many authors \cite{LO,agterberg_2,rad09,ber09,4e}. In this state, the gap function is real and oscillates sinusoidally.

The pure PDW representation of State II is less
 known \cite{agterberg_2}. In this state, From Eq.\ (\ref{mon}) we find that  the PDW order parameter satisfy
 $\varphi_A^*\varphi_B^*\varphi_C\varphi_D=-|\varphi_A\varphi_B\varphi_C\varphi_D|$, namely,
  $(\varphi_B,\varphi_A,\varphi_C,\varphi_D)\equiv(\varphi_{Qy},\varphi_{-Qy},\varphi_{Qx},\varphi_{-Qx})=\varphi_0(e^{i\phi_1},e^{i\phi_2},ie^{i\phi_3},ie^{i[\phi_2+\phi_2-\phi_3]})$ where the three $\phi_i$ are not fixed by the
 Free
 energy and represent the ${\rm U}(1)\times {\rm U}(1)\times {\rm U}(1)$ degeneracy of the ground state (we fix $\phi_i=0$ in the following). As shown in Fig.~\ref{PDW}, this PDW ground state is a vortex anti-vortex lattice phase which can be represented by the gap function $\Delta(\mathbf{x})=\Delta_0[\cos(Qx)+i\cos(Qy)]$. This gap function has position space zeroes at $Q\mathbf{x}=\pi(n+1/2,m+1/2)$ with integer $n,m$. Near these zeroes, the gap function becomes $\Delta(\mathbf{x})\approx \Delta_0Q(-1)^n[x+(-1)^{(n+m)}iy]$, explicitly showing the phase winding of the vortices and anti-vortices. This state breaks time-reversal symmetry continuously through the formation of the vortex anti-vortex lattice. This should be contrasted with the discrete breaking of time-reversal symmetry that appears in the CDW sector, once the theory extended beyond the hot spot approximation (see below). The ${\rm U}(1)\times {\rm U}(1)\times {\rm U}(1)$ degeneracy has a clear physical origin: one ${\rm U}(1)$ is associated with the usual SC phase and the other two ${\rm U}(1)$'s are associated with the acoustic phonons of the checkerboard lattice.  This ground state admits fractional vortices with one half the usual SC flux quantum \cite{agterberg_2}. Following Ref.\ \onlinecite{4e}, a treatment of thermal fluctuations associated with these vortices shows that the mean-field PDW order can in principle split into two transitions corresponding to the separation of the transition temperatures of the SC and the checkerboard order. Consequently, the high temperature phase transition can be into one of three possible states: the original mean-field PDW state, an orbital density wave state (with no SC phase coherence but with checkerboard order), or a spatially uniform charge-4e $d_{x^2-y^2}$ superconductor (this state is analogous to the spatially uniform charge-4e $s$-wave superconductor found in Ref.~\onlinecite{4e}, the $d_{x^2-y^2}$ symmetry follows from the relationship $\varphi_{Qx}\varphi_{-Qx}=-\varphi_{Qy}\varphi_{-Qy}$).
We note that a PDW state with the same symmetry has been proposed by P.A. Lee to account for the quasi-particle properties observed by ARPES measurements \cite{patrick}.

\subsection{Going beyond the hot spot approximation}
\label{5C}
In this subsection we neglect the difference in mean-field transition temperatures for CDW and PDW orders and consider instead what happens if we lift the restriction that $\rho_{i}$ and $\varphi_{i}$ couple only to specific hot spots in the Brillouin zone.
For definiteness, we constrain our discussions to a stripe state I, namely, consider a state with order parameters along bonds $A,B$ and $-A,-B$ (see Fig.\ \ref{bz}). For the stripe state, both $\Delta_A$ and $\Delta_B$ appear with the same magnitude, while their relative ${\rm SU}(2)$ ``phases" are arbitrary in the hot spot approximation.
  We assume that spin-mediated interaction has a finite ``width" in momentum space and allows some coupling between  the order parameters $\rho_A$ and $\varphi_A$ and fermions in the $B$ region and vice versa, and see how this breaks the symmetry between CDW and PDW components.

In the hot-spot approximation, the effective action for the stripe phase I is
\begin{align}
\mathcal{S}_{\rm eff}=&2\alpha (|\rho_A|^2+|\varphi_A|^2+|\rho_B|^2+|\varphi_B|^2)+\frac{1}{2}\beta\[(|\rho_A|^2+|\varphi_A|^2)^2+(|\rho_B|^2+|\varphi_B|^2)^2\].
\label{23}
\end{align}
The order parameter manifold is ${\rm SO}(4)\times {\rm SO}(4) \times {\mathbb Z}_2^{\rm rot}$, where ${\mathbb Z}_2^{\rm rot}$ corresponds to the choice of bond direction ($A,B$ or $C,D$) which is already made in Eq.\ (\ref{23}), and the two ${\rm SO}(4)$'s are for bonds $A$ and $B$ respectively. A pure CDW state or a pure PDW state are members of
  the ${\rm SO}(4)\times {\rm SO}(4) \times {\mathbb Z}_2^{\rm rot}$ manifold and for each of them the order parameter manifold is  ${\rm U}(1)\times {\rm U}(1) \times {\mathbb Z}_2^{\rm rot}$ where the two ${\rm U}(1)$'s are the phases  of $\rho(\varphi)_{A}$ and $\rho(\varphi)_B$ respectively.

We now go back to fermion-boson interaction term
    in the regions A and B
   and extend it to
\begin{align}
\mathcal{S}_{\rm int}=&\int d{k}~ \rho_A f_A(k)c^\dagger(k-Q/2)c(k+Q/2)+\rho_B f_B(k)c^\dagger(k-Q/2)c(k+Q/2)\nonumber\\
&+\int dk~ \varphi_Ag_A(k)c^\dagger(k+Q'/2)c^\dagger(-k+Q'/2)+\varphi_Bg_B(k)c^\dagger(k-Q'/2)c^\dagger(-k+Q'/2)+h.c.
\label{q}
\end{align}
 The form factors $f_{A(B)}(k)$ and $g_{A(B)}(k)$ are peaked around center-of-mass momentum of hot spots 1(5) and 2(6) and center-of-mass momentum of hot spots 3(7) and 4(8), but are
 no longer assumed to be $\delta-$functions of momenta.
 In Eq.\ (\ref{q}), $Q=Q_y$ and $Q'=-Q_x$ are the momenta of the CDW and PDW orders.  Integrating out fermions, we find that in this situation there appear
    non-zero couplings between $\rho_{A},\rho_{B}$, and $\varphi_{A},\varphi_{B}$.  The structure of the coupling terms is, however,
     different for CDW and PDW order parameters.

\begin{figure}
\includegraphics[width=0.45\columnwidth]{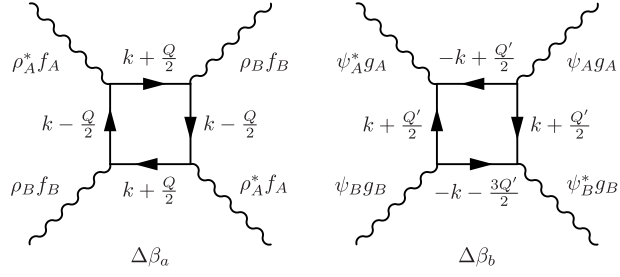}
\caption{Diagrammatic representations of the coefficients $\Delta\beta_a,\Delta\beta_b$.
We have associated form factors $f_{A,B}$ and $g_{A,B}$ to each vertex. The momenta are $Q = Q_y = (0,Q)$ and $Q' = -Q_x = (-Q,0)$.
 }
\label{db}
\end{figure}

 For CDW order parameter, integration over fermions yields a coupling term in the form
  \begin{align}
\Delta\mathcal{S}_a&=\Delta\beta_a[(\rho_A\rho_B^*)^2+(\rho_B\rho_A^*)^2],\label{95_1}
\end{align}
where
\begin{align}
\Delta\beta_a=-\int f_A^2(k)f_B^2(k)G^2(k-Q/2)G^2(k+Q/2).
\label{95}
\end{align}
We show the diagrammatic representation of $\Delta \beta_a$ in Fig.\ \ref{db}.
In the hot spot approximation, when $f_A$ and $f_B$ are $\delta$-functions peaked at different momenta, the integral in (\ref{95}) vanishes.
  Away from this limiting case, there is some overlap between $f_A$ and $f_B$ and the integral in (\ref{95}) is nonzero and
   Eq.\ (\ref{95_1}) breaks the degeneracy of the ${\rm U}(1) \times {\rm U}(1)$ manifold of CDW members of the stripe state I.

  The prefactor $\Delta\beta_a$
 has been evaluated in Sec.\ V\ C of Ref.\ \onlinecite{charge} for a particular model form of  $f_{A,B}$ and was found to be positive.
  Expressing $\rho_{A,B}$ as $\rho_{A,B}=|\rho_{A,B}|e^{i\phi_{A,B}}$, we obtain from Eq.\ (\ref{95_1})
  \begin{align}
  \Delta\mathcal{S}_a&=2\Delta\beta_a|\rho_A|^2|\rho_B|^2\cos[2(\phi_A-\phi_B)].\label{95_2}
  \end{align}
  For $\Delta\beta_a>0$ the phase difference $\phi_A-\phi_B$ is selected by (\ref{95_2})
  to be $\pm \pi/2$.  This lowers the ${\rm U}(1)\times {\rm U}(1)$ symmetry in the CDW sector to ${\rm U}(1)\times{\mathbb Z}_2$, where ${\mathbb Z}_2$ corresponds to two choices for the relative phase.
  Because under time-reversal $\rho_A\rightarrow \rho_B$  and, hence,  $\phi_A-\phi_B\rightarrow\phi_B-\phi_A$, the discrete ${\mathbb Z}_2$ symmetry is directly associated with the time-reversal and we represent it as  ${\mathbb Z}_2^{\rm trs}$ below.  Substituting $\phi_A-\phi_B =\pm \pi/2$ into (\ref{95_2}) we obtain
  \begin{align}
  \Delta\mathcal{S}_a&=- 2\Delta\beta_a|\rho_A|^2|\rho_B|^2 ,\label{95_2_2}
  \end{align}

On the other hand, for PDW order, the coupling terms which would depend on the phases of $\varphi_A$ and $\varphi_B$ are forbidden: as $\varphi_A$ and $\varphi_B$ carry opposite momenta, the term $(\varphi_A\varphi_B^*)^2$ cannot be present in the action as it would violate the momentum conservation. It has been shown~\cite{agterberg_2} that the only term which couples PDW components $\varphi_A$ and $\varphi_B$ is
\begin{align}
\Delta\mathcal{S}_b&=\Delta\beta_b|\varphi_A|^2|\varphi_B|^2,
\label{97}
\end{align}
where
\begin{align}
\Delta\beta_b=&\int g_A^2(k)g_B^2(k-Q'/2)G^2(k+Q'/2)G(-k+Q'/2)G(-k-{3Q'}/2).\label{99}
\end{align} We show the diagrammatic representation of $\Delta \beta_b$ in Fig.\ \ref{db}.

The presence of the additional terms given by  Eqs.\ (\ref{95_2_2}) and (\ref{97}) breaks the ${\rm SO}(4)$ symmetry of the effective action. To see this, we re-express $\rho$'s and $\varphi$'s  as
\begin{align}
\rho_A=&\|\Delta_A\|\cos\theta_Ae^{i\phi_A}{\rm,~~~and~~~}\varphi_A=\lVert\Delta_A\|\sin\theta_Ae^{i\varphi_A},\nonumber\\
\rho_B=&\|\Delta_B\|\cos\theta_Be^{i\phi_B}{\rm,~~~and~~~}\varphi_B=\|\Delta_B\|\sin\theta_Be^{i\varphi_B},
\label{1024}
\end{align}
where $\|\Delta_A\|\equiv \sqrt{|\rho_A|^2+|\varphi_A|^2}$,
and re-write the effective action as
\begin{align}
\mathcal{S}_{\rm eff}=&2\alpha(\|\Delta_A\|^2+\|\Delta_B\|^2)+\frac{\beta}{2}\[\|\Delta_A\|^4+\|\Delta_B\|^4\]\nonumber\\
&-\(2\Delta\beta_a\cos^2\theta_A\cos^2\theta_B-\Delta\beta_b\sin^2\theta_A\sin^2\theta_B\)\|\Delta_A\|^2\|\Delta_B\|^2.
\label{saturdaynight}
\end{align}
 Eq.\ (\ref{saturdaynight}) explicitly depends on $\theta_{A,B}$ and hence the ${\rm SO}(4)$ symmetry is broken.
 The outcome depends on the interplay between $\Delta\beta_a$ and $\Delta\beta_b$.
  Comparing the diagrams in Fig.\ \ref{db} we see that in the diagram for  $\Delta\beta_a$, all four fermions can be placed
   near the FS as there are only two momenta involved: $k + Q/2$ and $k-Q/2$, while in the the diagram for  $\Delta\beta_b$ one cannot do this
   as there are there different internal momenta there: $k+Q'/2$, $-k+Q'/2$, and $-k -3 Q'/2$ ($Q = Q_y$ and $Q' = -Q_x$),
   and  if two  are placed on the FS then the third has to be away from it.  As a result
   $\Delta\beta_a$ is much larger than $\Delta\beta_b$.

Using this, we immediately find from  Eq.\ (\ref{saturdaynight}) that the effective action is minimized when $\theta_A=\theta_B=0$.
    This means that  the extra terms in the action break O(4) symmetry between CDW and PDW in favor of CDW order (the symmetry is lowered
     from ${\rm SO}(4)\times {\rm SO}(4) \times {\mathbb Z}_2^{\rm rot}$ to ${\rm U}(1)\times {\mathbb Z}_2^{\rm trs}\times {\mathbb Z}_2^{\rm rot}$, where
 ${\rm U}(1)$ is the common phase of $\rho_A$ and $\rho_B$ and ${\mathbb Z}_2^{\rm trs}$ corresponds to two choices for the relative phase between $\rho_A$ and $\rho_B$).

\subsubsection{Properties of a pure CDW ground state}
\label{5D}

The properties of the pure CDW state have been studied before~\cite{charge,tsvelik,rahul} so we will be brief.
 The order parameter manifold for the stripe CDW phase with $|\rho_A| = |\rho_B| = \rho$ and $|\rho_C| = |\rho_D| = 0$
 is  ${\rm U}(1)\times {\mathbb Z}_2^{\rm trs}\times {\mathbb Z}_2^{\rm rot}$. The ${\rm U}(1)$ component is a common phase between $\rho_A$ and $\rho_B$ and its selection in the ordered CDW state
 reflects the breaking of a translational symmetry by an incommensurate CDW order.
   The order parameters which do not depend on the common phase but reflect the breaking of ${\mathbb Z}_2$ symmetries are composite order parameters
   $|\rho_A|^2 - |\rho_C|^2 = \pm \rho^2$ for  ${\mathbb Z}_2^{\rm rot}$ and  $\rho_A \rho^*_B = \pm i \rho^2$.
   The condition $\rho_A \rho^*_B = \pm i \rho^2$ together with $|\rho_A| = |\rho_B| = \rho$ implies that
   \beq
   \rho_A = \rho e^{\pm i \pi/4} e^{i\phi},~~~\rho_B = \rho e^{\mp i \pi/4} e^{i \phi}.
   \eeq
   Such an order has both charge modulation, given by $\rho_A + \rho_B$,  and  current modulation, given by $\rho_A -\rho_B$.  In the case we consider the
    charge modulation is along $y$ direction. Current modulation is also along $y$, but the current itself flows along $x$.
This order breaks time-reversal symmetry and also breaks mirror symmetries around $x$ and $y$ directions  (all three eigenfunctions change sign under
 ${\mathbb Z}_2^{\rm trs}$ which transforms $\rho_A \to \rho_B$).

 In a more general treatment than the one we presented above, a coupling of $\rho$ fields to fermions away from corresponding hot spots also leads
  to the modification of the quadratic terms in the effective action -- the quadratic form decouples between $\rho^{Q_y} \equiv \rho_A + \rho_B$ and $J^{Q_y} \equiv
  \rho_A - \rho_B$ (i.e., $\mathcal{S}_{\rm eff} = \alpha_{\rho} (\rho^{Q_y})^2  + \alpha_{J} (J^{Q_y})^2$ and $\alpha_\rho$ becomes negative at a higher $T$ than $\alpha_J$). In this situation,
$\rho^{Q_y}$ orders first and $J^{Q_y}$ acquires a non-zero condensation value  at a lower $T$ (with the phase difference $\pm \pi/2$ compared to
  $\rho^{Q_y}$).  The $ {\mathbb Z}_2^{\rm trs}$ symmetry gets broken only below a lower transition temperature.  At low $T$, both $\rho^{Q_y}$ and $J^{Q_y}$ have non-zero, but non-equal expectation values, i.e.,
   \begin{align}
   \rho_A &= \frac{1}{2} \left(\rho^{Q_y} + J^{Q_y} \right) =  \frac{1}{2} \left(|\rho^{Q_y}| \pm i |J^{Q_y}| \right) e^{i\phi},\nonumber\\
\rho_B &= \frac{1}{2} \left(\rho^{Q_y}- J^{Q_y} \right) = \frac{1}{2}  \left(|\rho^{Q_y}| \mp i |J^{Q_y}| \right) e^{i\phi}.
\label{0603}
   \end{align}

\subsection{Combining the effects of curvature and of coupling to fermions away from hot spots}
\label{5E}
We now return to our consideration of the effective action. Combining the effects of the curvature and of coupling of the order parameters $\rho$ and $\varphi$ to fermions away from hot spots, we obtain an  effective action in the form
\begin{align}
\mathcal{S}_{\rm eff}=&2\alpha_\rho (|\rho_A|^2+|\rho_B|^2) + 2 \alpha_\varphi (|\varphi_A|^2+|\varphi_B|^2)+\frac{1}{2}\beta\[(|\rho_A|^2+|\varphi_A|^2)^2+(|\rho_B|^2+|\varphi_B|^2)^2\] \nonumber\\
&-2 \Delta\beta_a|\rho_A|^2|\rho_B|^2 + \Delta\beta_b|\varphi_A|^2|\varphi_B|^2
\label{23_1}
\end{align}
where $\alpha_\rho(T)\sim(T-T_{\rm CDW})$ and $\alpha_\varphi(T)\sim(T-T_{\rm PDW})$ and $T_{\rm PDW} > T_{\rm CDW}$.
The last two terms in (\ref{23_1})  are from Eqs.\ (\ref{95_2}) and (\ref{97}), and we have set $\phi_A-\phi_B=\pm\pi/2$.

 To simplify the presentation, we set $\Delta\beta_b=0$ since it is much smaller than $\Delta\beta_a$.
  We also assume that the coupling to fermions away from hot spots is weak and set $2 \Delta\beta_a< \beta$. Then the action is positive-definite.

Eq.\ (\ref{23_1}) is symmetric with respect to $A\rightarrow B$ and there is no repulsion term between $A$ and $B$, hence we have for the ground state,
\begin{align}
|\rho_A|&=|\rho_B|\equiv\rho\nonumber\\
|\varphi_A|&=|\varphi_B|\equiv\varphi,
\end{align}
Using this, we rewrite the effective action as
\begin{align}
\mathcal{S}_{\rm eff}=&4\alpha_\rho\rho^2+4\alpha_\varphi\varphi^2+\beta(\rho^2+\varphi^2)^2-2\Delta\beta_a\rho^4.
\label{233_2}
\end{align}
The extremal values of Eq.\ (\ref{233_2}) are at
\begin{align}
\rho=0,~~~{\rm or}~~~&4\alpha_\rho+2\beta(\rho^2+\varphi^2)-4\Delta\beta_a\rho^2=0\nonumber\\
\varphi=0,~~~{\rm or}~~~&4\alpha_\varphi+2\beta(\rho^2+\varphi^2)=0.
\label{244}
\end{align}
Simple calculations show that solutions of Eq.\ (\ref{244}) are with either $\rho =0$ or $\varphi =0$, hence
the CDW order and PDW order do not coexist.

At $T_{\rm CDW}<T<T_{\rm PDW}$, a pure PDW state develops, with $\varphi=\sqrt{{-2\alpha_\varphi}/{\beta}}$. At lower temperatures $T<T_{\rm CDW}<T_{\rm PDW}$, a pure CDW state  also becomes possible. For this CDW state we obtain $\rho=\sqrt{{-2\alpha_\rho}/{(\beta-2\Delta\beta_a)}}$.

To decide  whether CDW or PDW state is more favorable at a low $T$ we compare the values of the effective action for pure PDW and CDW states:
\begin{align}
\mathcal{S}_{\rm PDW}&=\frac{-4\alpha_\varphi^2}{\beta}\label{ccc}\\
\mathcal{S}_{\rm CDW}&=\frac{-4\alpha_\rho^2}{\beta-2\Delta\beta_a}.\label{ppp}
\end{align}
 If the difference between $\alpha_\varphi$ and $\alpha_\rho$ is small, CDW order definitely wins (we recall that $\Delta\beta_a>0$).
In this situation, the system undergoes a first-order transition from PDW to CDW state at some $T < T_{\rm CDW}$.  If the difference
 between $\alpha_\varphi$ and $\alpha_\rho$ is larger, the PDW order may survive down to $T=0$.

\subsection{Going beyond mean-field approximation}
\label{5F}
For the action given by Eq.\ (\ref{23_1}),  the three symmetries associated with CDW order: ${\rm U}(1)$, ${\mathbb Z}_2^{\rm trs}$ and ${\mathbb Z}_2^{\rm rot}$,
 get broken at the same temperature $T_{\rm CDW}$. This, however, is true only within the  mean-field approximation.
  Once we go beyond mean-field theory and include fluctuation effects, discrete symmetries (in our case ${\mathbb Z}_2^{\rm trs}$ and ${\mathbb Z}_2^{\rm rot}$) get broken at higher temperatures  than the temperature $T_{\rm CDW}$ at which the continuous ${\rm U}(1)$ symmetry gets broken~\cite{starykh, arun_1,arun_2, fernandes,nie,charge,tsvelik}.
 The two $\mathbb{Z}_2$ symmetries do not generally get broken at the same temeperature; which one is higher depends on the relative strength of the corresponding symmetry breaking terms in the action (for a detailed discussion, see Sec.\ VI of Ref.\ \onlinecite{charge}).
In the intermediate temperature range  $\langle\rho_{\rm A,B,C,D}\rangle=0$ but $\langle|\rho_{\rm A,B}|^2-|\rho_{\rm C,D}|^2\rangle\neq0$ and/or $i\langle\rho_A\rho_B^*-\rho_A^*\rho_B\rangle\neq0$. Previous studies, originally done for Fe-pnictides~\cite{fernandes} and then extended to the spin-fermion model, which we study here~\cite{charge,tsvelik},  have found that the feedback from discrete symmetry breaking pushes
 the onset temperature $T_{\rm CDW}$ for the primary CDW order to a higher value.

This effect also exists for PDW order as the corresponding order parameter manifold  ${\rm U}(1)\times {\rm U}(1)\times {\mathbb Z}_2^{\rm rot}$ also contains
 a ${\mathbb Z}_2$ component which beyond mean-field orders at a higher $T$ than mean-field $T_{\rm PDW}$ and pushes the onset temperature for
   ${\rm U}(1) \times {\rm U}(1)$ breaking to a higher $T$.
We expect such an enhancement of onset temperature to be weaker than that for CDW order, because for the latter there are two discrete $Z_2$ degrees of freedom in the order parameter manifold, and each ordering of $Z_2$ degree of freedom increases the susceptibility of the primary field and hence increases the onset temperature for the breaking of the corresponding $U(1)$ degree of freedom.
    If this effect overshoots the difference between $\alpha_\rho$ and $\alpha_\varphi$ in Eq.\ (\ref{23_1}) then the system only develops a stripe CDW order and no stripe PDW order.  The  actual calculation of the transition temperatures for CDW and PDW orders in the presence of preemptive ${\mathbb Z}_2$ orders requires
    going well beyond mean-field and is beyond the scope of current work.

 In principle, it is possible that out of the two continuous ${\rm U}(1)$ symmetries for PDW, which can be viewed as
 one  translational and one gauge symmetry, one ${\rm U}(1)$ symmetry is broken prior to the other. One such proposal is a charge-4e superconductor~\cite{4e}, which is a bound state of two PDW order parameters with opposite momenta, which corresponds to $\langle\varphi_A\varphi_B\rangle$ in our case (for analogous proposal for magnetic systems see Ref. \onlinecite{starykh}). In such a state ${\rm U}(1)$ gauge symmetry is broken while ${\rm U}(1)$ translational symmetry is preserved.  This may also lift the onset temperature
  for the primary PDW order parameter (i.e., the temperature below which both ${\rm U}(1)$ symmetries are broken).
 However, whether this happens in the spin-fermion model is beyond the scope of this work.

\section{Secondary orders induced by CDW and PDW in a checkerboard state}
\label{sec:6}

We now return to mean-field theory and consider state II with checkerboard order in which CDW/PDW orders develop along both horizontal and vertical bonds.  We assume that the conditions are such that at low $T$ both CDW and PDW components of the order parameter are present. It is known that the presence of multiple order parameters  can induce ``secondary" orders through third order coupling terms. For example, PDW orders of momenta $(0,\pm Q)$ alone are known to give rise to CDW orders of momenta $(0,\pm 2Q)$~\cite{patrick,agterberg, kivelson}. Here we examine a possibility that a simultaneous presence of CDW/PDW orders induces a
 homogeneous charge-2e superconducting order \cite{ber09,agterberg_2,patrick}.

\subsection{General Ginzburg-Landau Theory}
\label{6A}
 Prior to examining the microscopic theory, we present a general symmetry based analysis.
 In State II CDW and PDW order parameters carry momenta $\pm Q_x=(\pm Q,0)$ and $\pm Q_y=(0,\pm Q)$.
 Like in Eq.\ (\ref{2}), we introduce CDW and PDW order parameters as
   \begin{align}
 i\sigma_{\alpha\beta}\varphi_{Q_x}^{k}\sim c_\alpha(k-Q_x/2) c_\beta(-k-Q_x/2){\rm,~~~and~~~}\delta_{\alpha\beta}\rho_k^{Q_x}\sim c_\alpha^\dagger(k+Q_x/2) c_\beta(k-Q_x/2),\nonumber\\
  i\sigma_{\alpha\beta}\varphi_{Q_y}^{k}\sim c_\alpha(k-Q_y/2) c_\beta(-k-Q_y/2){\rm,~~~and~~~}\delta_{\alpha\beta}\rho_k^{Q_y}\sim c_\alpha^\dagger(k+Q_y/2) c_\beta(k-Q_y/2).
  \label{friday}
 \end{align}
Recall that the momenta carried by $\varphi_{Q}^k$ and by $\rho_k^Q$ are both $Q$, independent on whether the order is CDW or PDW.

Following the discussion at the end of subsection \ref{5D} we
split the $k$-dependence of $\rho_k^Q$
and generalize Eq.\ (\ref{0603}) by including form factors $f_{1,2}$ to
\begin{align}
\rho_k^{Q_x}&=\[\rho^{Q_x} f_1^{x}(k)+ J^{Q_x} f_2^{x}(k)\]/2 \nonumber\\
\rho_k^{Q_y}&=\[\rho^{Q_y} f_1^{y}(k)+J^{Q_y} f_2^{y}(k)\]/2,
 \end{align}
where for $\rho_k^{Q_x}$, ${\bf k}$ is along the $y$ direction in the hot spot model (and close to it in a generic model),
 and for $\rho_k^{Q_y}$, ${\bf k}$ is along the $x$ direction. The form-factor $f_1^{x} (k)$ is even under mirror reflection $k_y\rightarrow -k_y$, and $f_2^x(k)$ is odd,
 and $f_1^{y} (k)$ and $f_2^y(k)$ are even/odd  under $k_x\rightarrow -k_x$.
 We remind that in real space $\rho^{Q_{x(y)}}$ corresponds to a  charge density modulation with ordering momenta $Q_{x(y)}$, and $J^{Q_{x(y)}}$ corresponds to a bond current density modulation flowing in the $y(x)$ direction with ordering momenta $Q_{x,y}$.

  One can easily verify that the order parameters $\rho^{Q_x}$ and $J^{Q_x}$ belong to different irreducible representations of the little group $G_{Q_x}=\{E,C_{2x},\sigma_y,\sigma_z\}$ which consists of the set of rotation elements that keep $Q_x$ unchanged:
   $\rho^{Q_x}$ belongs to the $A_1$ representation as it does not change under applications of all elements of $G_{Q_x}$,  while $J^{Q_x}$ belongs to the $B_1$ representation -- it is odd under $\sigma_y$ and $C_{2x}$ and even under $\sigma_z$.

  The PDW order parameter $\varphi_{Q_x}^{k}$ is,  by definition,  even under $\sigma_y: k_y\rightarrow -k_y$ because it is  spin-singlet.
 We then define
 \begin{align}
 \varphi_{Q_x}^{k}\equiv \varphi_{Q_x}g_1^{x}(k),
 \end{align}
 where ${\bf k}$ in $g_1^x(k)$ is predominantly along $y$ and $g_1^x(k)$  is even under $k_y\rightarrow -k_y$.  One can easily verify that the
  order parameter $\varphi_{Q_x}$ belongs to the $A_1$ representation of the group $G_{Q_x}$.

We also note that  $\rho_k^{-Q}=(\rho_k^Q)^*$, while $\varphi_Q^k$ and $\varphi_{-Q}^k$ are generally different order parameters.
We therefore use
 $\rho^{Q_{x,y}}$, $J^{Q_{x,y}}$, $\varphi_{Q_{x,y}}$, and  $\varphi_{-Q_{x,y}}$ as independent order parameters.
    The properties of these order parameters under a $C_4$ lattice rotation ($x\rightarrow y$, $y\rightarrow -x$) and time-reversal are summarized in  Table I.

   \begin{table}
 \caption{The transformation of order parameters $\rho^{Q_{x,y}}$, $J^{Q_{x,y}}$, $\varphi_{Q_{x,y}}$, and  $\varphi_{-Q_{x,y}}$ under $C_4$ lattice rotation ($x\rightarrow y, y\rightarrow -x$) and time-reversal (TR). Particularly, we note that $J^{Q_{x}}$ transforms into $-J^{Q_y}$ under $C_4$. The minus sign is because $J^{Q_{x}}$ corresponds to a current which flows in $y$ direction, and this current should flow in $-x$ direction after a $C_4$ lattice rotation.}
 \begin{ruledtabular}
 \begin{tabular}{ccccc}
 Original OP& $\rho^{Q_{x}}$&$J^{Q_{x}}$&$\varphi_{Q_{x}}$&$\varphi_{-Q_{x}}$ \\
Under $C_4$ & $\rho^{Q_{y}}$&$-J^{Q_{y}}$&$\varphi_{Q_{y}}$&$\varphi_{-Q_{y}}$ \\
Under TR & $\rho^{Q_{x}}$&$-J^{Q_{x}}$&$(\varphi_{-Q_{x}})^*$&$(\varphi_{Q_{x}})^*$ \\\hline
Original OP& $\rho^{Q_{y}}$&$J^{Q_{y}}$&$\varphi_{Q_{y}}$&$\varphi_{-Q_{y}}$ \\
Under $C_4$ & $\rho^{-Q_{x}}\equiv(\rho^{Q_{x}})^*$&$J^{-Q_{x}}\equiv(J^{Q_{x}})^*$&$\varphi_{-Q_{x}}$&$\varphi_{Q_{x}}$\\
Under TR & $\rho^{Q_{y}}$&$-J^{Q_{y}}$&$(\varphi_{-Q_{y}})^*$&$(\varphi_{Q_{y}})^*$
    \end{tabular}
 \end{ruledtabular}
 \label{table:1}
 \end{table}
 Because
  $\varphi_{Q}$,
   $\rho^{Q}$, and $J^{Q}$
   all carry momentum $Q$,
    it is possible to construct the translational invariant products of the form $\rho^Q\varphi_{-Q}$, $(\rho^Q)^*\varphi_{Q}$, and $J^{Q}\varphi_{-Q}$, $(J^{Q})^*\varphi_{Q}$. These products have the same symmetry properties as homogeneous charge-2e SC order. The fact that $\rho^Q$ and $J^Q$ belong to different representations of $G_Q$ implies that the homogeneous SC order parameters that are proportional to $\rho^Q$ and $J^Q$ also belong to different representations.

    Consider separately the couplings between $\rho^Q$ and $\varphi_{Q}$ and between $J^Q$ and $\varphi_{Q}$.
The couplings between $\rho^Q$ and $\varphi_{Q}$ allow for four different types of homogeneous SC orders: $\rho^{Q_x}\varphi_{-Q_x},(\rho^{Q_x})^*\varphi_{Q_x},\rho^{Q_y}\varphi_{-Q_y}$, and $(\rho^{Q_y})^*\varphi_{Q_y}$. In our theory, only even parity SC homogeneous order appears (spin-singlet pairing), so we will not consider the two odd-parity SC states. The  two even parity SC orders can be combined into
 \begin{align}
 \rho^{Q_x}\varphi_{-Q_x}+(\rho^{Q_x})^*\varphi_{Q_x}+\rho^{Q_y}\varphi_{-Q_y}+(\rho^{Q_y})^*\varphi_{Q_y} {,~~~{\rm and}~~~}
 \rho^{Q_x}\varphi_{-Q_x}+(\rho^{Q_x})^*\varphi_{Q_x}-\rho^{Q_y}\varphi_{-Q_y}-(\rho^{Q_y})^*\varphi_{Q_y}.
\end{align}
The first combination has $A_{1g}$ ($s$) symmetry and the second one has
 $B_{1g}$ ($d_{x^2-y^2}$) symmetry. Accordingly, we introduce $s$-wave and $d_{x^2-y^2}$-wave SC order parameters via
\begin{align}
c_\alpha(k)c_\beta(-k)\sim i\sigma_{\alpha\beta}^y[\Phi_s h_s(k)+\Phi_{d_{x^2-y^2}} h_{d_{x^2-y^2}}(k)],
\end{align}
where the form factors $h_s$ ($h_{d_{x^2-y^2}}$) are even (odd) under a $C_4$ lattice rotation, and write the Free energy to quadratic order in $\Phi_s$ and $\Phi_{d_{x^2-y^2}}$ as
\begin{align}
\mathcal{S}_{\Phi}=&\alpha_s|\Phi_s|^2+ \epsilon_s \{\Phi_s^*[\rho^{Q_x}\varphi_{-Q_x}+(\rho^{Q_x})^*\varphi_{Q_x}+\rho^{Q_y}\varphi_{-Q_y}+(\rho^{Q_y})^*\varphi_{Q_y}]+h.c.\}\nonumber\\
&+\alpha_{d_{x^2-y^2}}|\Phi_{d_{x^2-y^2}}|^2 +\epsilon_d\{\Phi_{d_{x^2-y^2}}^*[\rho^{Q_x}\varphi_{-Q_x}+(\rho^{Q_x})^*\varphi_{Q_x}-\rho^{Q_y}\varphi_{-Q_y}-(\rho^{Q_y})^*\varphi_{Q_y}]+h.c.\}.
\label{fridaynight}
\end{align}
 One can directly verify using Table I that Eq.\ (\ref{fridaynight}) is invariant under lattice $C_4$ rotation and time-reversal. The effective ``triple coupling" constants $\epsilon_s$ and $\epsilon_{d_{x^2-y^2}}$   can be expressed as the convolutions of fermionic Green's functions with form factors $h_s(k),f_1(k)g_1(k)$ and $h_{d_{x^2-y^2}}(k),f_1(k),g_1(k)$, respectively, and their values depend on the details of the underlying microscopic model (see next Subsection).

Minimizing Eq.\ (\ref{fridaynight}) with respect to $\Phi_s^*$ and $\Phi_{d_{x^2-y^2}}^*$ we obtain
\begin{align}
\Phi_s&=-\frac{\epsilon_s}{\alpha_s}[\rho^{Q_x}\varphi_{-Q_x}+(\rho^{Q_x})^*\varphi_{Q_x}+\rho^{Q_y}\varphi_{-Q_y}+(\rho^{Q_y})^*\varphi_{Q_y}]\nonumber\\
\Phi_{d_{x^2-y^2}}&=-\frac{\epsilon_{d_{x^2-y^2}}}{\alpha_{d_{x^2-y^2}}}[\rho^{Q_x}\varphi_{-Q_x}+(\rho^{Q_x})^*\varphi_{Q_x}-\rho^{Q_y}\varphi_{-Q_y}-(\rho^{Q_y})^*\varphi_{Q_y}].
\label{n_10}
\end{align}

 Consider next the coupling between $J^Q$ and $\varphi_{Q}$. The same analysis as we did for the previous case
 shows that the two relevant bilinear combinations of $J^Q$ and $\varphi_{-Q}$ are
 \begin{align}
J^{Q_x}\varphi_{-Q_x}-(J^{Q_x})^*\varphi_{Q_x}+J^{Q_y}\varphi_{-Q_y}-(J^{Q_y})^*\varphi_{Q_y}{,~~~{\rm and}~~~}
J^{Q_x}\varphi_{-Q_x}-(J^{Q_x})^*\varphi_{Q_x}-J^{Q_y}\varphi_{-Q_y}+(J^{Q_y})^*\varphi_{Q_y}.
\end{align}
The first combination has $B_{2g}$ ($d_{xy}$) symmetry and the second one has
 $A_{2g}$ ($d_{xy} \times d_{x^2-y^2}$) symmetry. Accordingly, we introduce $d_{xy}$ and $A_{2g}$ SC order parameters via
\begin{align}
c_\alpha(k)c_\beta(-k)\sim i\sigma_{\alpha\beta}^y[\Phi_{d_{xy}} h_{d_{xy}}(k)+\Phi_{{A_{2g}}} h_{A_{2g}}(k)].
\end{align}
where the form factors $h_{d_{xy}}$ ($h_{A_{2g}}$) are odd (even) under a $C_4$ lattice rotation and write
the Free energy for these two orders as
\begin{align}
\mathcal{S}_{\Phi}=&\alpha_s|\Phi_{d_{xy}}|^2+ \epsilon_{d_{xy}} \{\Phi_{d_{xy}}^*[J^{Q_x}\varphi_{-Q_x}-(J^{Q_x})^*\varphi_{Q_x}+J^{Q_y}\varphi_{-Q_y}-(J^{Q_y})^*\varphi_{Q_y}]+h.c.\}\nonumber\\
&+\alpha_{d_{A_{2g}}}|\Phi_{A_{2g}}|^2 +\epsilon_{A_{2g}}\{\Phi_{A_{2g}}^*[J^{Q_x}\varphi_{-Q_x}-(J^{Q_x})^*\varphi_{Q_x}-J^{Q_y}\varphi_{-Q_y}+(J^{Q_y})^*\varphi_{Q_y}]+h.c.\}.
\label{fridaynightt}
\end{align}
Again, it can be directly verified from Table I that Eq.\ (\ref{fridaynightt}) is invariant under lattice $C_4$ rotation and time-reversal. The effective coupling constants $\epsilon_{d_{xy}}$ and $\epsilon_{A_{2g}}$  can be expressed as the convolution of fermionic Green's functions with form factors $h_{d_{xy}}(k),f_2(k)g_1(k)$ and $h_{A_{2g}}(k),f_2(k),g_1(k)$, respectively.

Minimizing Eq.\ (\ref{fridaynightt}) with respect to $\Phi_{d_{xy}}^*$ and $\Phi_{A_{2g}}^*$ we obtain
\begin{align}
\Phi_{d_{xy}}&=-\frac{\epsilon_{d_{xy}}}{\alpha_{d_{xy}}}[J^{Q_x}\varphi_{-Q_x}-(J^{Q_x})^*\varphi_{Q_x}+J^{Q_y}\varphi_{-Q_y}-(J^{Q_y})^*\varphi_{Q_y}]\nonumber\\
\Phi_{A_{2g}}&=-\frac{\epsilon_{A_{2g}}}{\alpha_{A_{2g}}}[J^{Q_x}\varphi_{-Q_x}-(J^{Q_x})^*\varphi_{Q_x}-J^{Q_y}\varphi_{-Q_y}+(J^{Q_y})^*\varphi_{Q_y}].
\end{align}

\subsection{Computation of the
triple
couplings within spin-fermion model}
\label{6B}

From a pure symmetry point of view, a homogeneous charge 2e SC order parameters with $s$, $d_{x^2-y^2}$, $d_{xy}$, and $A_{2g}$ symmetries  all emerge as secondary orders in a state in which CDW and PDW condensates are simultaneously present. In this subsection we
 evaluate the coefficients within our spin-fermion model and show that they vanish if we use linearized dispersion near hot spots
  but are non-zero when we keep the curvature of the FS $\kappa$  non-zero.
  If we only treat the curvature to leading order, i.e., neglect the curvature-induced difference between CDW and PDW orders, we find that
 only $s$ and $d_{xy}$ secondary SC orders develop. Beyond the leading order in $\kappa$,
 the other two secondary SC orders ($d_{x^2-y^2}$ and $A_{2g}$)  also likely emerge.

 To be specific, we consider a member of state II for which CDW order develops along one bond direction, say ($A,B$), and PDW order develops along the other bond direction ($C,D$).
 Such a ``orthogonal" configuration maximized the gain of energy due to the development of the secondary SC order and by this reason is a strong candidate for the
  actual CDW/PDW configuration in the state II ~\cite{coex}.

The
PDW/CDW order with CDW along ($A,B$) and PDW along ($C,D$)
is described as
\begin{align}
\Delta_A^{\mu\nu}=\(\begin{array}{cc}
0&\rho_A^*\\
-\rho_A&0
\end{array}\),~
\Delta_B^{\mu\nu}=\(\begin{array}{cc}
0&\rho_B\\
-\rho_B^*&0
\end{array}\),~
\Delta_C^{\mu\nu}=\(\begin{array}{cc}
\varphi_C&0\\
0&\varphi_C^*
\end{array}\),~\Delta_D^{\mu\nu}=\(\begin{array}{cc}
\varphi_D&0\\
0&\varphi_D^*
\end{array}\).
\label{da}
\end{align}
with
\begin{align}
|\rho_A|=|\rho_B|=|\varphi_C|=|\varphi_D|.
\label{460}
\end{align}
For such order, the constraint on the orientations of $\Delta_{A,B,C,D}^{\mu\nu}$, which, we remind, is
 $\Tr(U_A^\dagger U_B U_C^\dagger U_D)=-2$ or Eq.\ (\ref{mon}),  becomes
\begin{align}
\rho_A^*\rho_B^*\varphi_C^*\varphi_D=|\rho_A\rho_B\varphi_C\varphi_D|.
\label{conf}
\end{align}

\begin{figure}
\begin{tabular}{cc}
\parbox{0.4\columnwidth}{\includegraphics[width=0.4\columnwidth]{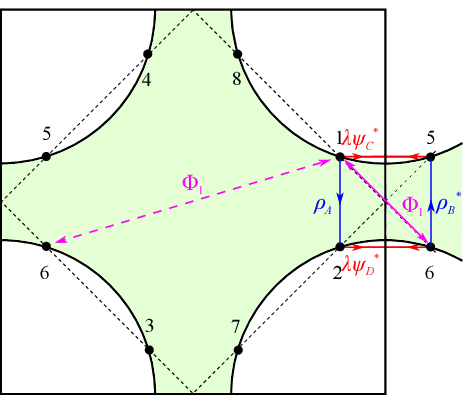}\\(a)}~~&
~~\parbox{0.4\columnwidth}{\includegraphics[width=0.4\columnwidth]{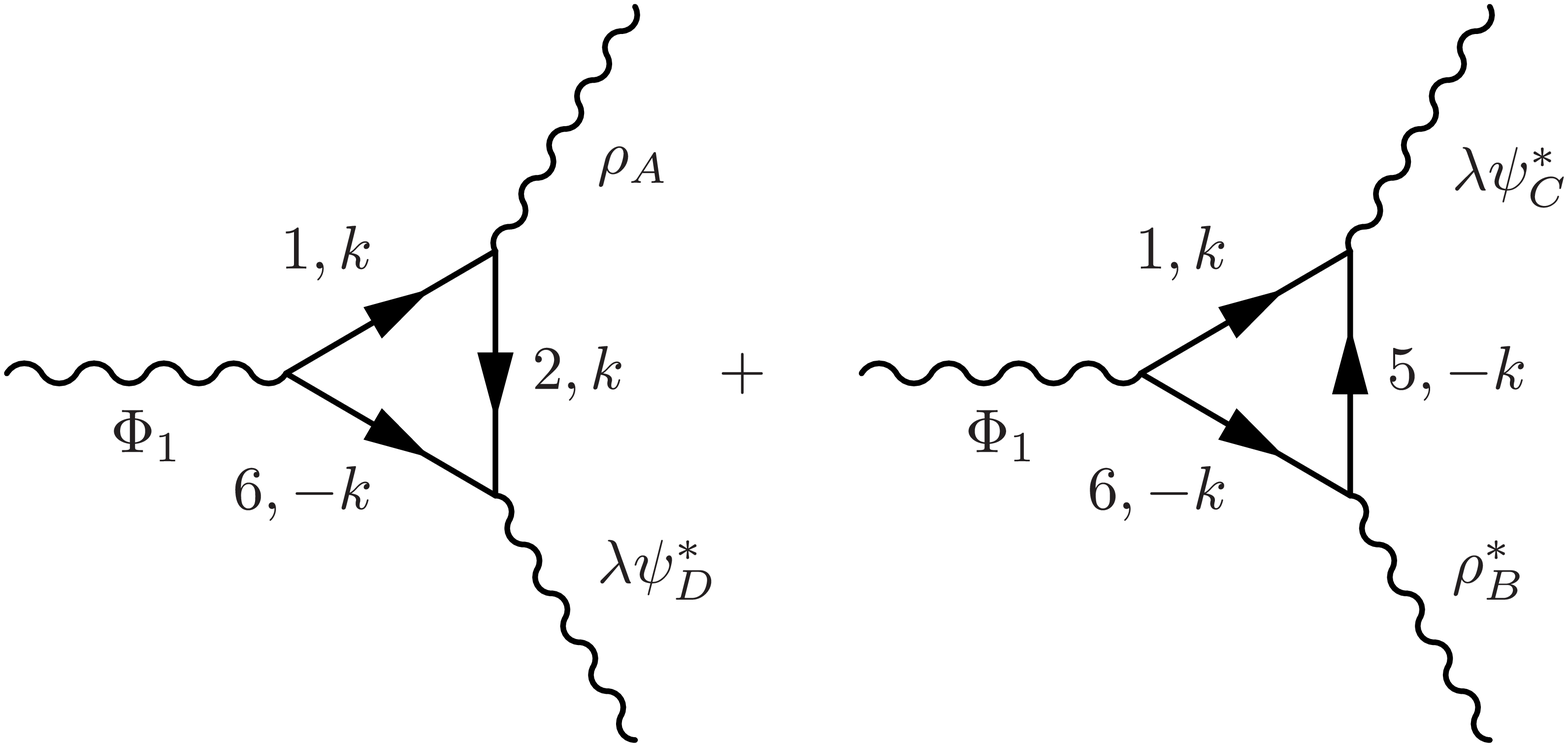}\\(b)\\\includegraphics[width=0.4\columnwidth]{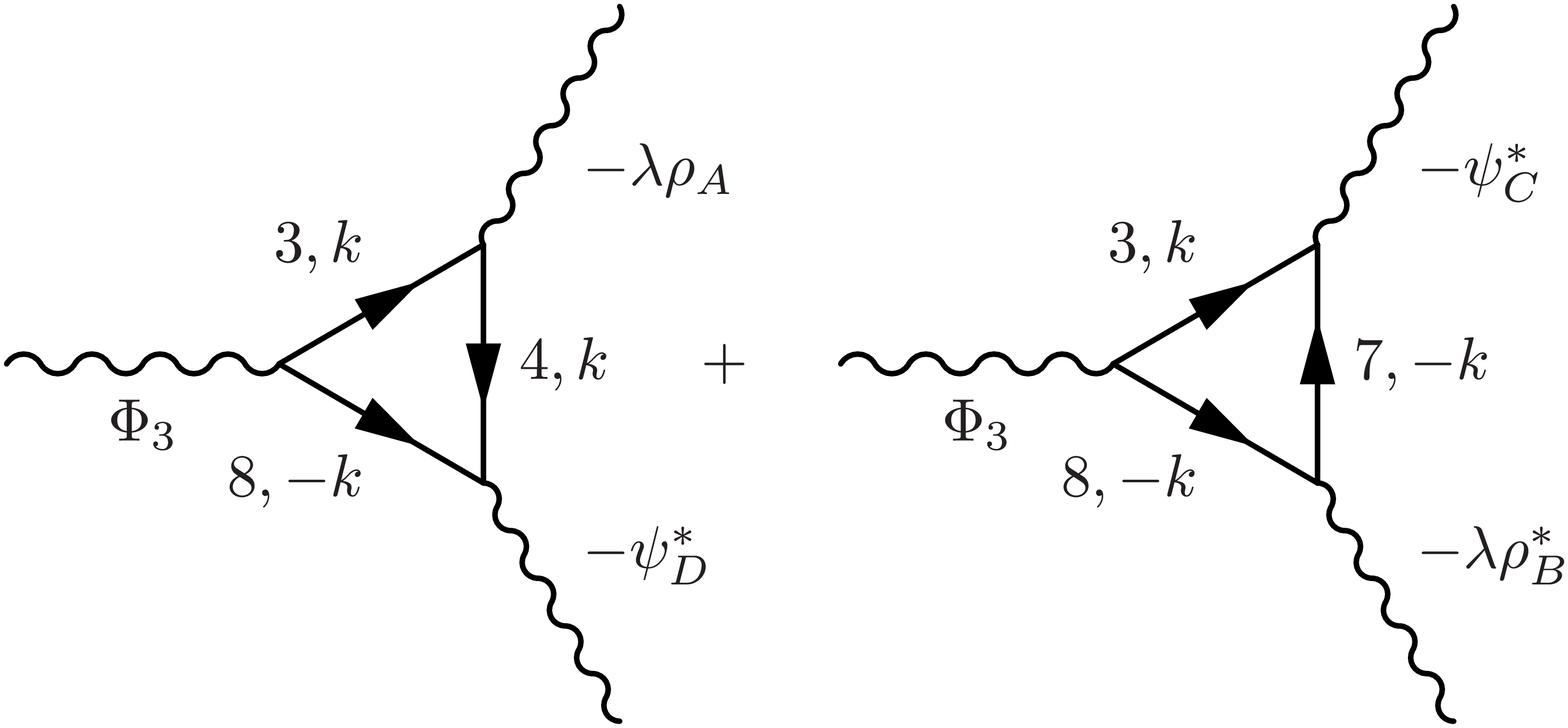}\\(c)}
\end{tabular}
\caption{Panel (a): A superconducting order parameter $\Phi_1$ between hot spots 1 and 6. We label CDW order $\rho$ with a line with an arrow, and PDW and homogeneous superconducting order with lines with double arrows. Panel (c): diagrams for coupling of $\Phi_1$ with $\rho$ and $\varphi$. Panel (b): diagrams for coupling of $\Phi_1$ with $\rho$ and $\varphi$. Comparing with Panel (b), the minus sign between CDW's and between PDW's cancel out.}
\label{bz2}
\end{figure}

Following the consideration in the previous Subsection, we define the homogeneous SC order parameter in terms of hot fermions as
\begin{align}
i\sigma^y_{\alpha\beta}\Phi_1c_{1\alpha}^\dagger(\tilde k)c_{6\beta}^\dagger(-\tilde k)+i\sigma^y_{\alpha\beta}\Phi_2c^\dagger_{2\alpha}(\tilde k)c^\dagger_{5\beta}(-\tilde k)+i\sigma^y_{\alpha\beta}\Phi_3 c^\dagger_{3\alpha}(\tilde k)c^\dagger_{8\beta}(-\tilde k)+i\sigma^y_{\alpha\beta}\Phi_4 c_{4\alpha}^\dagger(\tilde k)c_{7\beta}^\dagger(-\tilde k).
\end{align}
Particularly, in Fig.\ \ref{bz2}(a), we show $\Phi_1$ between hot spots 1 and 6 and its relations to CDW order $\rho$ and PDW order $\varphi$ at hot spots, and we show its diagrammatic representation in Fig.\ \ref{bz2}(b). From Fig.\ \ref{bz2}(b) and Eqs.\ (\ref{sd},\ref{da}) we express the triple coupling term involving $\Phi_1$, $\rho$, and $\varphi$ as,
 \begin{align}
 \mathcal{S}_{\Phi_1\rho\varphi}=-2\lambda \Phi_1( Y_{126}\rho_A\varphi_D^*+ Y_{156}\rho_B^*\varphi_C^*)+h.c.,
 \label{sp}
 \end{align}
 where
 \begin{align}
 Y_{126}=&\int d\tilde k G_1(\tilde k)G_2(\tilde k)G_6(-\tilde k),\nonumber\\
 Y_{156}=&\int d\tilde k G_1(\tilde k)G_5(-\tilde k)G_6(-\tilde k),
 \label{n_9}
 \end{align}
 and the coefficient $(-2)$ comes from spin summation, $\tilde k=(\omega_m, \bf \tilde k)$, and $\bf \tilde k$ is the momemtum deviation from a hot spot.
 The coefficients $Y_{126}$ and $Y_{156}$ are equal
 by symmetry
 because the two integrals in (\ref{n_9})
 are related by inversion: $1\rightarrow 6$, $2\rightarrow5$ and $\tilde k\rightarrow -\tilde k$. To get a finite value $Y_{126}=Y_{156}$, one, however,
  has to keep the curvature of the Fermi surface~\cite{23}, otherwise $Y_{126}=Y_{156}$ would be zero.  In Eq. (\ref{sp}) we included the curvature into
   the Green's functions in (\ref{n_9}) but otherwise assumed that both CDW and PDW order parameters change by the same $-\lambda$ once we change the momentum
   ${\bf k}$  by $\pi$.

  One can write the same triple coupling for other pairs of hot spots. We obtain
 \begin{align}
  \mathcal{S}_{\Phi_2\rho\varphi}=-2\lambda \Phi_2( Y_{256}\rho_B\varphi_D^*+ Y_{125}\rho_A^*\varphi_C^*)+h.c.,\nonumber\\
 \mathcal{S}_{\Phi_3\rho\varphi}=-2\lambda \Phi_3( Y_{348}\rho_A\varphi_D^*+ Y_{378}\rho_B^*\varphi_C^*)+h.c.,\nonumber\\
 \mathcal{S}_{\Phi_4\rho\varphi}=-2\lambda \Phi_4( Y_{478}\rho_B\varphi_D^*+ Y_{347}\rho_A^*\varphi_C^*)+h.c.,
 \end{align}
 where
  \begin{align}
 Y_{125}=&\int G_1(k)G_2(k)G_5(-k),~~Y_{256}=\int G_1(k)G_5(-k)G_6(-k),\nonumber\\
 Y_{348}=&\int G_3(k)G_4(k)G_8(-k),~~Y_{378}=\int G_3(k)G_7(-k)G_8(-k),\nonumber\\
 Y_{347}=&\int G_3(k)G_4(k)G_7(-k),~~Y_{478}=\int G_4(k)G_7(-k)G_8(-k).
 \end{align}
  The corresponding diagrams for $\Phi_3$ are shown in Fig.\ \ref{bz2}(c).

 We verified that all $Y$ terms are equal, i.e., $Y_{126}=Y_{156}=Y_{125}=Y_{256}=Y_{348}=Y_{378}=Y_{347}=Y_{478}\equiv Y$.

\begin{figure}
\centerline{\parbox{0.35\columnwidth}{\includegraphics[width=0.35\columnwidth]{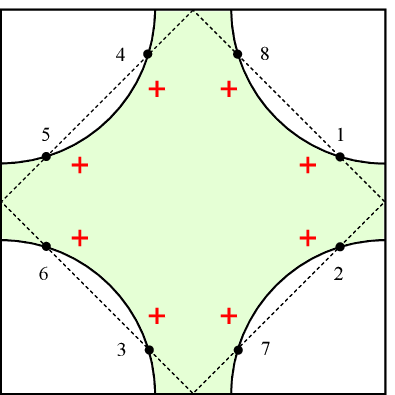}\\(a)}~~~~
\parbox{0.35\columnwidth}{\includegraphics[width=0.35\columnwidth]{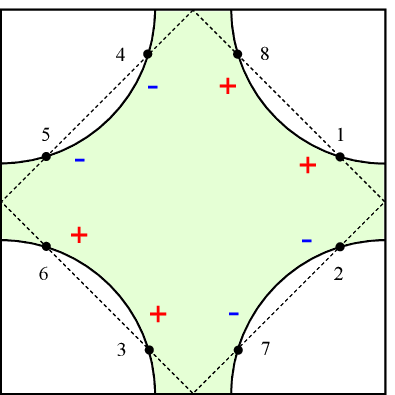}\\(b)}}
\caption{The signs of order parameter at hot spots 1-8 for $s$-wave SC (Panel (a)) and $d_{xy}$-wave SC (Panel (b)).}
\label{sdxy}
\end{figure}

  As a result,  the effective action
becomes
 \begin{align}
   \mathcal{S}_{\Phi}=&\alpha_\phi{[|\Phi_1|^2+|\Phi_2|^2+|\Phi_3|^2+|\Phi_4|^2]}\nonumber\\
   &-2\lambda Y [\Phi_1( \rho_A\varphi_D^*+ \rho_B^*\varphi_C^*)+\Phi_2( \rho_B\varphi_D^*+\rho_A^*\varphi_C^*)+\Phi_3(\rho_A\varphi_D^*+\rho_B^*\varphi_C^*)+\Phi_4( \rho_B\varphi_D^*+\rho_A^*\varphi_C^*)]+h.c.
  \label{500}
\end{align}
We see that triple coupling terms involving $\Phi_1$ and $\Phi_3$ and the ones involving $\Phi_2$ and $\Phi_4$ are identical.
 Comparing panels (b) and (c), we see that this equivalence is the direct consequence of the fact that both
$\rho$ and $\varphi$ change by $-\lambda$ under the momentum transformation by $(\pi,\pi)$ (i.e., under transformation from the FS region 1256 to the region 3478
 in  Fig.\ \ref{bz2}(a).
 In this situation, the prefactors for $\Phi_1$ and $\Phi_3$ and for $\Phi_2$ and $\Phi_4$ become equivalent when  $\rho$ and $\varphi$ terms are
  combined in the three-leg diagrams.
Minimizing the action in Eq.\ (\ref{500}), we immediately obtain that ${\Phi_1}=\Phi_3$ and $\Phi_2=\Phi_4$. Recalling the positions of hot spots 1, 2, 3, and 4  this condition only allows for $s$ and $d_{xy}$ symmetry as both $d_{x^2-y^2}$ and $A_{2g}$ SC order would require ${\Phi_1}=-\Phi_3$ (see Fig.\ \ref{sdxy}).
  In other words, to leading order in the curvature,
 $\epsilon_{d_{x^2-y^2}}$ and $\epsilon_{A_{2g}}$ in Eqs. (\ref{fridaynight}) and (\ref{fridaynightt}) are zero.

Once we include into our consideration the fact that the curvature $\kappa$ also breaks the symmetry between CDW and PDW orders, the ratios between CDW and PDW
 order parameters under momentum transformation by $(\pi,\pi)$, e.g., $\rho_{-A}/\rho_A$ and $\varphi_{-A}/\varphi_{A}$, do not have to be the same, and from Eq.\ (\ref{forty-one}), we have in general
  $\lambda_\rho = \rho_{-A}/\rho_A$ and $\lambda_\varphi = \varphi_{-A}/\varphi_{A}$ with $\lambda_\varphi - \lambda_\rho \propto \kappa^2$.
  In this case, we verified that $\Phi_1$ and $\Phi_3$ are not identical and, as a result, $\epsilon_{d_{x^2-y^2}}$ and $\epsilon_{A_{2g}}$ are nonzero
   However, the magnitudes of  $\epsilon_{d_{x^2-y^2}}$ and $\epsilon_{A_{2g}}$ contain extra $\kappa^2$ compared to $\epsilon_s$ and $\epsilon_{d_{xy}}$, respectively.
   On the other hand, $\alpha_\phi$ term in (\ref{500}) likely favor $d_{x^2-y^2}$ superconductivity, once we go beyond hot spot approximation, so for not very small $\kappa$ a secondary SC instability in $d_{x^2-y^2}$ channel is a possibility.

Let's continue the analysis to leading order in $\kappa$ when $s$ and $d_{xy}$  SC orders develop. The issue we now address is what is the relative phase between these two secondary orders.
We  show that there is a $\pi/2$ phase difference between $s$ component and $d_{xy}$ component, i.e., the pairing symmetry is $s+id_{xy}$.  From Eqs.\ (\ref{460}) and (\ref{conf}), we find $\rho_A\varphi_D^*=\rho_B^*\varphi_C^*$, and $\rho_B\varphi_D^*=\rho_A^*\varphi_C^*$. Defining $\Phi_s\equiv \Phi_1+\Phi_2+\Phi_3+\Phi_4=2(\Phi_1+\Phi_2)$ and $\Phi_{d_{xy}}\equiv\Phi_1-\Phi_2+\Phi_3-\Phi_4=2(\Phi_1-\Phi_2)$, we can rewrite Eq.\ (\ref{500}) as
\begin{align}
 \mathcal{S}_{\Phi}=&\alpha_s|\Phi_s|^2+\alpha_d|\Phi_{d_{xy}}|^2-\lambda Y [\Phi_s(\rho_A+\rho_B)\varphi_D^*+\Phi_{d_{xy}}(\rho_A- \rho_B)\varphi_D^*)]+h.c.\nonumber\\
\equiv&\alpha_s|\Phi_s|^2+\alpha_d|\Phi_{d_{xy}}|^2-\lambda Y [\Phi_s\rho_{AB}\varphi_D^*+i\Phi_{d_{xy}}J_{AB}\varphi_D^*)]+h.c.
  \label{501}
\end{align}
where $\alpha_s=\alpha_d=\alpha_\phi/8$, and
in the last line we have defined $\rho_{AB}$ and $J_{AB}$ through $\rho_{A}\equiv(\rho_{AB}+iJ_{AB})/2$ and $\rho_{B}\equiv(\rho_{AB}-iJ_{AB})/2$.
 As $|\rho_A|=|\rho_B|$, $\rho_{AB}$ and $J_{AB}$ must have the same phase, hence $J_{AB}=r \rho_{AB}$ and $r$ is real. Using this relation, we
  re-write (\ref{501}) as
\begin{align}
 \mathcal{S}_{\Phi}=&\alpha_s|\Phi_s|^2+\alpha_d|\Phi_{d_{xy}}|^2-\lambda Y (\Phi_s+ir\Phi_{d_{xy}})\rho_{AB}\varphi_D^*+h.c.
\end{align}
From this action we clearly see that the induced SC order should have a $(s+id_{xy})$-wave symmetry. In the spin-fermion model, $\alpha_s=\alpha_d$, and whether $s$ or $d$ component is dominant depends on the value of $r$.
Going beyond spin-fermion model, generally we have $\alpha_s>\alpha_d$ since on-site (Hubbard) interaction which is independent on momenta in $k$-space strongly surpresses $s$-wave but not $d$-wave. On the other hand, if time-reversal symmetry or mirror symmetry is preserved, then $\rho_A=\rho_B$, since they transform to each other under these operations. In this case $J_{AB}=0$, hence the induced SC can only be $s$-wave. This is no surprise since coexistence of $s$ and $id_{xy}$ breaks time-reversal symmetry and mirror symmetry.

We briefly consider the induced SC order if the checkerboard state is a generic one, with CDW and PDW components along all bonds. In this case, the CDW/PDW order parameters are given by the general form Eq.\ (\ref{1106}).
By the same reasoning, the secondary homogeneous SC order is induced by CDW components along one bond direction ($A,B$ or $C,D$) and PDW components along the other bond direction ($C,D$ or $A,B$). Following the same procedure as before, we find
 \begin{align}
    \mathcal{S}_{\Phi}=&\alpha_\phi{[|\Phi_1|^2+|\Phi_2|^2+|\Phi_3|^2+|\Phi_4|^2]}\nonumber\\
&-2\lambda Y [\Phi_1( \rho_A\varphi_D^*+\rho_D^*\varphi_A^*+ \rho_B^*\varphi_C^*+\rho_C\varphi_B^*)+\Phi_2( \rho_B\varphi_D^*+\rho_D\varphi_B^*+\rho_A^*\varphi_C^*+\rho_C^*\varphi_A^*)\nonumber\\
  &+\Phi_3( \rho_A\varphi_D^*+\rho_D^*\varphi_A^*+ \rho_B^*\varphi_C^*+\rho_C\varphi_B^*)+\Phi_4( \rho_B\varphi_D^*+\rho_D\varphi_B^*+\rho_A^*\varphi_C^*+\rho_C^*\varphi_A^*)]+h.c.
  \label{502}
\end{align}
Once again we see  that $\Phi_1=\Phi_3$ and $\Phi_2=\Phi_4$, hence the induced SC should be a mixture of $s$-wave and $d_{xy}$-wave only ($d_{x^2-y^2}$-wave and $A_{2g}$-wave do not occur
as long as we assume that $\rho_{-A}/\rho_A = \varphi_{-A}/\varphi_A$, etc)
 However, in this generic case the relative phase of $s$-component and $d_{xy}$-component is not set to be $\pm \pi/2$.

\section{Conclusion and application to the cuprates}

In this paper we studied the interplay between PDW and CDW orders within the spin-fermion model.
The model was originally put forward to account for $d$-wave superconductivity near the onset of magnetism, but over the last few years it has been realized that it
describes not only a homogeneous $d$-wave superconductivity but also charge orders, such as bond order with momentum $(Q,Q)$ and CDW order with momentum $(Q,0)/(0,Q)$.
In this work, we have shown that the model also describes PDW -- a pair-density-wave superconducting order with a non-zero total momentum of the pair.
 We have demonstrated that the
 (approximate) ${\rm SU}(2)$ particle-hole symmetry of the spin-fermion model, previously used to link a homogeneous $d$-wave superconductivity and charge bond order,
  also links a CDW order and PDW order which in this regard become  intertwined orders.  Keeping the ${\rm SU}(2)$ symmetry explicit, we found that PDW and CDW order parameters can be combined into a larger PDW/CDW order parameter $\Delta$. The PDW/CDW order parameter is bilinear in ${\rm SU}(2)$-symmetric fermions and has ${\rm SU}(2)\times {\rm SU}(2) \sim {\rm SO}(4)$ symmetry. We developed a covariant Ginzburg-Landau theory for four PDW/CDW order parameters $\Delta_{A,B,C,D}$, and studied the ground state configurations. Depending on parameters, we have found two possible ground states: a ``stripe" state, where either $\Delta_{A,B}$ or $\Delta_{C,D}$ orders, and a ``checkerboard" state, where all four order parameters $\Delta_{A,B,C,D}$ develop.  We
   showed that the ${\rm SO}(4)$ symmetry between CDW and PDW can be broken by two separate effects already within mean-field theory. One is the inclusion of Fermi surface curvature, which  selects a PDW order immediately below the instability temperature.
  Another is the overlap between different hot regions, which
   favors CDW order at low temperatures. We showed that, for the stripe state,  the  competition between the two effects
    gives rise to first-order transition from PDW to CDW inside the ordered state.  We argued that beyond mean-field, the critical temperature for CDW order is additionally increased compared to that for PDW order due to feedback from the breaking of an extra ${\mathbb Z}_2$ time-reversal symmetry in the CDW state.
    If this additional increase overshoots the effect of Fermi surface curvature, the system only develops a CDW order.
      For the checkerboard state, we considered a situation  when both CDW and PDW orders are present at low $T$ and showed that the presence of both condensates  induces a secondary composite order with $s+id_{xy}$ symmetry.  This order further lowers the energy of state II.

  State II, in which both CDW and PDW are present, is our proposed candidate for the charge-ordered state in underdoped cuprates. The gain of energy due to the
   secondary SC order is maximized for the "orthogonal" state in which CDW order develops between a pair of hot spots along, say, vertical direction and PDW order
 develops between a pair of hot spots along horizontal direction (or vise versa). One can easily make sure that in such configuration CDW and PDW order parameters
  actually carry the {\it same} momenta.  Despite belonging to a checkerboard state in our classification, it has
   all features of stripe CDW order. Namely, it breaks $C_4$ lattice rotational symmetry and $Z_2$ time-reversal symmetry. At the same time, the presence of the PDW component allows one to explain quantitatively~\cite{coex} ARPES data in the pseudogap state~\cite{shen_a}. Without a PDW component, one could explain ARPES data
   for the cuts  near Brillouin zone boundary~\cite{charge}, but not closer to zone diagonal~\cite{patrick}.

Another issue relevant to the physics of the cuprates is the interplay between our PDW/CDW order and d-wave superconductivity.  In the present paper we
 have restricted the analysis to temperatures above $T_c$.  The extension of the present  work to $T < T_c$ shows~\cite{coex} that secondary SC order
  induced by CDW/PDW and d-wave SC order couple below $T_c$ in such a way that the measured SC gap becomes nodeless.  We propose to do careful ARPES measurements
    of the SC gap in the whole co-existence region with the charge order to verify this claim.

\begin{acknowledgments}
We thank   E.\ Abrahams,  W.\ A.\ Atkinson, E.\ Berg,  G.\ Y.\ Cho, D.\ Chowdhury,
 R.\ Fernandes, E.\ Fradkin,  M.\ Greven,  S.\ Kivelson, P.\ A.\ Lee, S.\ Lederer,  C.\ P\'epin, S.\ Raghu,  S.\ Sachdev, D.\ Scalapino, J.\ Schmalian, L.\ Taillefer, A.\ Tsvelik and S.\ Vishveshwara
 for fruitful discussions. The work was supported by the DOE grant DE-FG02-ER46900 (AC and YW) and by NSF grant No. DMR-1335215 (DFA).
 \end{acknowledgments}

 \appendix
 \section{Doping dependence of $T_{\rm CDW}$ and $T_{\rm PDW}$ via variation of the chemical potential}\label{ratio}

 Within the spin-fermion model the doping $x$ comes into play via two effects: through the variation of the magnetic correlation length $\xi$ and through the variation of the chemical potential $\mu$. In the main text we assumed that the latter effect is small and considered the variation of $T_{\rm CDW}$ and $T_{\rm PDW}$ with magnetic $\xi$.  Here we present quantitative study how $T_{\rm CDW}$ and $T_{\rm PDW}$ vary upon changing $\mu$. To be brief, we consider $SU(2)$-symmetric hot spot model in which $T_{\rm CDW} = T_{\rm PDW}$.

 The variation of the chemical potential $\mu$ changes the location of hot spots and, accordingly, the ratio of Fermi velocities $v_y/v_x$. The latter
   determines $\sqrt{S_1S_2}$ and $T_{\rm CDW} \sim ge^{-1/\sqrt{S_1S_2}}$ (see Eqs (\ref{7_1}) and (\ref{ac_last}) and Eq.\ (17) in Ref.\ \onlinecite{charge}).

\begin{figure}[h]
\includegraphics[width=0.5\columnwidth]{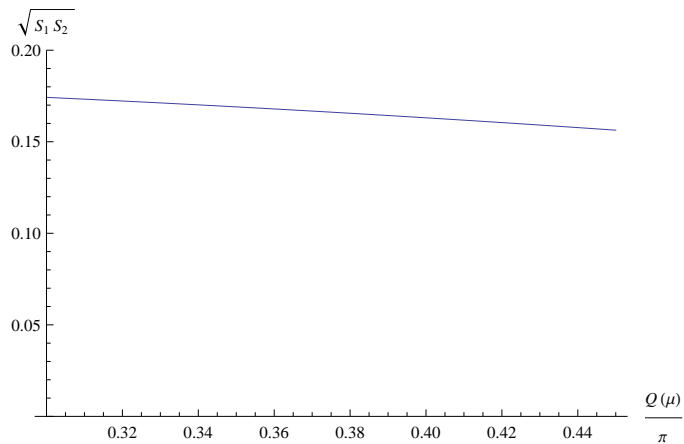}
\caption{the dependence of parameter $\sqrt{S_1S_2}$ on chemical potential $\mu$. We vary $\mu$ such that the CDW ordering momentum $Q$ varies in the range $(0.3\pi, 0.45\pi)$, as measured by experiments.}
\label{vxvy}
\end{figure}

 As experimental input, we use the dispersion  in nearly optimally doped Pb-Bi2201 from Ref.\ \onlinecite{shen_a}: $\epsilon(k_x, k_y)=-2t(\cos{k_x}+\cos{k_y})-4t^{'} (\cos{k_x} \cos{k_y})-2 t^{''}
(\cos{2k_x}+\cos{2k_y})-4 t^{'''}(\cos{2k_x} \cos{k_y}+\cos{k_x} \cos{2k_y})-\mu$, with
$t =0.22 {\rm eV}$, $t^{'} = -0.034315 {\rm eV}$, $t^{''} = 0.035977 {\rm eV}$, $t^{'''} = -0.0071637 {\rm eV}$.  We keep $\mu$ as a variable to account for different dopings. Depending on doping, the CDW wave-vector for this material ranges from  $Q=0.3\pi$ (optimally doped) to $Q=0.45\pi$ (underdoped, see Ref.\ \onlinecite{hudson}).  We  vary the chemical potential $\mu$ to match the variation of  $Q(\mu)$ in the range $(0.3\pi, 0.45\pi)$. For this  range, we find that the parameter $\sqrt{S_1S_2}$ is essentially a constant (see Fig.\ {\ref{vxvy}}). We therefore  conclude that, at least in the range relevant to cuprates, the effect of varying $\mu$ on the transition temperatures $T_{\rm CDW}$ and $T_{\rm PDW}$ is very small and can be neglected.

 \section{Proof of the ${\rm SO}(4)\times {\rm SO}(4)$ symmetry of the PDW/CDW Ginzburg-Landau action}
 \label{app:1}
 In this Appendix we show explicitly that the PDW/CDW action Eq.\ (\ref{seff}) has an ${\rm SO}(4)\times {\rm SO}(4)$ symmetry.

 It is helpful to re-express each of the four $SU(2)$ phases $U_{A,B,C,D}$ via four real variables
 \begin{align}
 &U_A^{\mu\nu}=\(\begin{array}{cc}
A_1+iA_2&A_3-iA_4\\
-A_3-iA_4&A_1-iA_2
\end{array}\),~
U_B^{\mu\nu}=\(\begin{array}{cc}
B_1+iB_2&B_3-iB_4\\
-B_3-iB_4&B_1-iB_2
\end{array}\),\nonumber\\
&U_C^{\mu\nu}=\(\begin{array}{cc}
C_1+iC_2&C_3-iC_4\\
-C_3-iC_4&C_1-iC_2
\end{array}\),~
U_D^{\mu\nu}=\(\begin{array}{cc}
D_1+iD_2&D_3-iD_4\\
-D_3-iD_4&D_1-iD_2
\end{array}\).
 \end{align}
 It is easy to verify that $\sqrt{A_1^2+A_2^2+A_3^2+A_4^2}=1$, and the same relation holds for $B$'s, $C$'s and $D$'s.

We can then re-express the last term of Eq.\ (\ref{seff}) as
\begin{align}
\mathcal{S}_{c}=&2\lambda^2I_4\sqrt{(|\rho_A|^2+|\varphi_A|^2)(|\rho_B|^2+|\varphi_B|^2)(|\rho_C|^2+|\varphi_C|^2)(|\rho_D|^2+|\varphi_D|^2)}\[\Tr(U_A^\dagger U_CU_B^\dagger U_D)+h.c.\]\nonumber\\
=&8\lambda^2I_4\sqrt{(|\rho_A|^2+|\varphi_A|^2)(|\rho_B|^2+|\varphi_B|^2)(|\rho_C|^2+|\varphi_C|^2)(|\rho_D|^2+|\varphi_D|^2)}\nonumber\\
&\times\[(A_1C_1+A_2C_2+A_3C_3+A_4C_4)(B_1D_1+B_2D_2+B_3D_3+B_4D_4)\right.\nonumber\\
&-(A_1C_2-A_2C_1+A_3C_4-A_4C_3)(B_1D_2-B_2D_1-B_3D_4+B_4D_3)\nonumber\\
&-(A_1C_3-A_2C_4-A_3C_1+A_4C_2)(B_1D_3+B_2D_4-B_3D_1-B_4D_2)\nonumber\\
&\left.+(A_1C_4+A_2C_3-A_3C_2-A_4C_1)(B_1D_4-B_2D_3+B_3D_2-B_4D_1)\].\nonumber\\
=&8\lambda^2I_4\sqrt{(|\rho_A|^2+|\varphi_A|^2)(|\rho_B|^2+|\varphi_B|^2)(|\rho_C|^2+|\varphi_C|^2)(|\rho_D|^2+|\varphi_D|^2)}\nonumber\\
&\times\(\begin{array}{cccc}A_1& A_2& A_3& A_4\end{array}\)\(\begin{array}{cccc}
C_1&-C_2&-C_3&-C_4\\
C_2&C_1&C_4&-C_3\\
C_3&-C_4&C_1&C_2\\
C_4&C_3&-C_2&C_1
\end{array}\)
\(\begin{array}{cccc}
D_1&-D_2&-D_3&D_4\\
D_2&D_1&D_4&D_3\\
D_3&-D_4&D_1&-D_2\\
-D_4&-D_3&D_2&D_1
\end{array}\)
\(\begin{array}{c}
B_1\\
-B_2\\
-B_3\\
B4
\end{array}\).
\label{555}
\end{align}

Recalling that $\sqrt{C_1^2+C_2^2+C_3^2+C_4^2}=1$ and $\sqrt{D_1^2+D_2^2+D_3^2+D_4^2}=1$, one can easily verify that the matrix product
\begin{align}
S_{CD}\equiv\(\begin{array}{cccc}
C_1&-C_2&-C_3&-C_4\\
C_2&C_1&C_4&-C_3\\
C_3&-C_4&C_1&C_2\\
C_4&C_3&-C_2&C_1
\end{array}\)
\(\begin{array}{cccc}
D_1&-D_2&-D_3&D_4\\
D_2&D_1&D_4&D_3\\
D_3&-D_4&D_1&-D_2\\
-D_4&-D_3&D_2&D_1
\end{array}\)
\end{align}
 is an ${\rm SO}(4)$ matrix. In fact, it is known mathematically~\cite{elfrinkhof} that every SO(4) matrix can be {\it uniquely} decomposed into such a matrix product. The matrix composed of $C$'s in this decomposition represents a left-isoclinic rotation in four-dimensional Euclidean space, and the matrix composed of $D$'s represents a right-isoclinic rotation (note the difference in their matrix structures).

 We define four-dimensional vectors $V_A=(A_1,A_2,A_3,A_4)$ and $V_B=(B_1,-B_2,-B_3,B_4)$ and re-write Eq. (\ref{555}) as
 \begin{align}
 \mathcal{S}_c=&8\lambda^2I_4\sqrt{(|\rho_A|^2+|\varphi_A|^2)(|\rho_B|^2+|\varphi_B|^2)(|\rho_C|^2+|\varphi_C|^2)(|\rho_D|^2+|\varphi_D|^2)}\times V_A^{i}S_{CD}^{ij}V_B^j
 \label{556}
 \end{align}
 where sum over $i,j$ from 1 to 4 is assumed.

   Eq.\ (\ref{556}) is invariant under two SO(4) rotations, represented by $S_A$ and $S_B$,
 \beq
 V_A\rightarrow V_A'\equiv S_AV_A,~~~V_B\rightarrow V_B'\equiv S_BV_B,~~~{\rm and}~~~S_{CD}\rightarrow S_{CD}'\equiv S_AS_{CD}S_B^T.
 \label{a5}
 \eeq
 The matrix $S_{CD}'$ is also an $SO(4)$ matrix, which means it can be uniquely decomposed as
 \begin{align}
S'_{CD}\equiv S_AS_{CD}S_B^T=\(\begin{array}{cccc}
C'_1&-C'_2&-C'_3&-C'_4\\
C'_2&C'_1&C'_4&-C'_3\\
C'_3&-C'_4&C'_1&C'_2\\
C'_4&C'_3&-C'_2&C'_1
\label{a6}
\end{array}\)
\(\begin{array}{cccc}
D'_1&-D'_2&-D'_3&D'_4\\
D'_2&D'_1&D'_4&D'_3\\
D'_3&-D'_4&D'_1&-D'_2\\
-D'_4&-D'_3&D'_2&D'_1
\end{array}\).
\end{align}
 This in turn implies that $S_A$ and $S_B$ uniquely determine the transformations of $C$'s and $D$'s.
 We see that from Eqs.\ (\ref{a5}) and (\ref{a6}) that the symmetry of $\mathcal{S}_{c}$ is $\rm SO(4)\times SO(4)$.
  It is easy to show that all other terms in the effective action of Eq.\ (\ref{seff})  are also invariant under these two SO(4) transformations. Therefore, the full continuous symmetry of the effective action (\ref{seff}) is ${\rm SO}(4)\times {\rm SO}(4)$.

  \section{The evaluation of Eqs. (\ref{tr5})-(\ref{tr8})}
  \label{curv}
For Eqs.\ (\ref{tr5}) and (\ref{tr6}), we introduce
 $z=\tilde k_y/\sqrt{\gamma|\omega_m|}$ and use the zero-temperature form of $\Sigma(\omega_m)\approx(2/3)\sgn\sqrt{\omega_0\omega_m}$, and rewrite Eqs.\ (\ref{tr5},\ref{tr6}) as
\begin{align}
\varphi_{A}=&-\frac{3T_{\rm PDW}}{8}\sum_{m}\frac{\varphi_{-A}}{|\omega_m|}\int_{0}^{\infty}\frac{dz}{\sqrt{z^2+1}}\frac{1}{\[\sqrt{z^2+1}+\(1+\sqrt{9|\omega_m|/4\omega_0}\)/4\]^2}\nonumber\\
&+\frac{3\kappa^2T_{\rm PDW}}{8}\sum_{m}\frac{\varphi_{-A}}{|\omega_m|}\int_{0}^{\infty}\frac{dz}{\sqrt{z^2+1}}\frac{\gamma|\omega_m|z^4/k_F^2}{\[\sqrt{z^2+1}+\(1+\sqrt{9|\omega_m|/4\omega_0}\)/4\]^4}\label{tr17}\\
\varphi_{-A}=&=\frac{3T_{\rm PDW}}{8}\sum_{m}\frac{\varphi_{A}}{|\omega_m|}\int_{0}^{\infty}\frac{dz}{\sqrt{z^2+1}}\frac{1}{\(1+\sqrt{9|\omega_m|/4\omega_0}\)^2/16+z^2}\nonumber\\
&-\frac{3\kappa^2T_{\rm PDW}}{8}\sum_{m}\frac{\varphi_{A}}{|\omega_m|}\int_{0}^{\infty}\frac{dz}{\sqrt{z^2+1}}\frac{4z^2{\gamma|\omega_m|}(z^2+1)^2/k_F^2}{\[\(1+\sqrt{9|\omega_m|/4\omega_0}\)^2/16+z^2\]^3}\label{tr18}.
\end{align}
We have only kept the leading order dependence on $\kappa$. For Eqs.\ (\ref{tr7}) and (\ref{tr8}), we have after identical calculations
\begin{align}
\rho_{A}=&-\frac{3T_{\rm CDW}}{8}\sum_{m}\frac{\rho_{-A}}{|\omega_m|}\int_{0}^{\infty}\frac{dz}{\sqrt{z^2+1}}\frac{1}{\[\sqrt{z^2+1}+\(1+\sqrt{9|\omega_m|/4\omega_0}\)/4\]^2}\nonumber\\
&+\frac{9\kappa^2T_{\rm CDW}}{8}\sum_{m}\frac{\rho_{-A}}{|\omega_m|}\int_{0}^{\infty}\frac{dz}{\sqrt{z^2+1}}\frac{\gamma|\omega_m|z^4/k_F^2}{\[\sqrt{z^2+1}+\(1+\sqrt{9|\omega_m|/4\omega_0}\)/4\]^4}\label{tr9}\\
\rho_{-A}=&-\frac{3T_{\rm CDW}}{8}\sum_{m}\frac{\rho_{A}}{|\omega_m|}\int_{0}^{\infty}\frac{dz}{\sqrt{z^2+1}}\frac{1}{\(1+\sqrt{9|\omega_m|/4\omega_0}\)^2/16+z^2}\nonumber\\
&+\frac{3\kappa^2T_{\rm CDW}}{8}\sum_{m}\frac{\rho_{A}}{|\omega_m|}\int_{0}^{\infty}\frac{dz}{\sqrt{z^2+1}}\frac{\(1+\sqrt{9|\omega_m|/4\omega_0}\)^2\kappa^2{\gamma|\omega_m|}(z^2+1)^2/(4k_F^2)}{\[\(1+\sqrt{9|\omega_m|/4\omega_0}\)^2/16+z^2\]^3}\label{tr0}.
\end{align}
Evaluating Eqs.\ (\ref{tr17},\ref{tr18},\ref{tr9},\ref{tr0}) we obtain Eqs. (\ref{forty-one}) in the main text.

\end{document}